\documentclass[fleqn,usenatbib,useAMS]{mnras}

\usepackage{graphicx}	\usepackage{amsmath}	\usepackage{amssymb}	\usepackage{multicol}        \usepackage{bm}		\usepackage{pdflscape}	\usepackage{booktabs}
\usepackage{rotating}
\usepackage{tabulary}

\newcommand{\kms}{\,km\,s$^{-1}$} \newcommand{\Ha}{H$\alpha$} \newcommand{\imfit}{\textsc{imfit}} \newcommand{\Msun}{M$_{\odot}$}
\newcommand{\sizeratio}{$r_{\mathrm{e}, {\rm{H}\alpha}}/r_{\mathrm{e}, R_{\mathrm{c} } }$}
\defcitealias{Tiley:2020}{Paper 1}

\usepackage[T1]{fontenc}
\usepackage{ae,aecompl}

\usepackage{newtxtext,newtxmath}

\title[Environmental impacts on cluster galaxies]{K-CLASH: Strangulation and Ram Pressure Stripping in Galaxy Cluster Members at 0.3 < $z$ < 0.6}

\author[S. P. Vaughan et al.]{Sam P. Vaughan,$^{1, 2, 3}$\thanks{Contact e-mail: \href{mailto:sam.vaughan@sydney.edu.au}{sam.vaughan@sydney.edu.au} (SPV) }
Alfred L. Tiley,$^{4, 5}$ Roger L. Davies,$^{3}$ Laura J. Prichard,$^{6}$
\newauthor Scott M. Croom,$^{1,2}$ Martin Bureau,$^3$ John P. Stott,$^{7}$ Andrew Bunker,$^{3}$  Michele Cappellari,$^3$ 
\newauthor  Behzad Ansarinejad$^{4}$  and Matt J. Jarvis$^{3,8}$ \\
$^{1}$Sydney Institute for Astronomy, School of Physics, Building A28, The University of Sydney, NSW 2006, Australia\\
$^{2}$ARC Centre of Excellence for All Sky Astrophysics in 3 Dimensions (ASTRO3D), Australia\\
$^{3}$Sub-department of Astrophysics, Department of Physics, University of Oxford, Denys Wilkinson Building, Keble Road, Oxford OX1 3RH, UK\\
$^{4}$International Centre for Radio Astronomy Research, The University of Western Australia, 35 Stirling Highway, Crawley WA 6009, Australia\\
$^{5}$Centre for Extragalactic Astronomy, Department of Physics, Durham University, South Road, Durham DH1 3LE, UK\\
$^{6}$Space Telescope Science Institute, 3700 San Martin Drive, Baltimore MD 21218, USA\\
$^{7}$Department of Physics, Lancaster University, Bailrigg, Lancaster LA1 4YB, UK\\
$^{8}$Department of Physics \& Astronomy, University of the Western Cape, Private Bag X17, Bellville, Cape Town, 7535, South Africa\\
}

\date{Accepted 2020 June 22. Received 2020 June 21; in original form 2019 November 18.}

\pubyear{2019}

\begin{document}
\label{firstpage}
\pagerange{\pageref{firstpage}--\pageref{lastpage}}
\maketitle

\begin{abstract}
Galaxy clusters have long been theorised to quench the star-formation of their members. This study uses integral-field unit observations from the $K$-band Multi-Object Spectrograph (KMOS) - Cluster Lensing And Supernova survey with \textit{Hubble} (CLASH) survey (K-CLASH) to search for evidence of quenching in massive galaxy clusters at redshifts $0.3<z<0.6$. We first construct mass-matched samples of exclusively star-forming cluster and field galaxies, then investigate the spatial extent of their H$\alpha$ emission and study their interstellar medium conditions using emission line ratios.  The average ratio of \Ha{} half-light radius to optical half-light radius (\sizeratio{}) for all galaxies is $1.14\pm0.06$, showing that star formation is taking place throughout stellar discs at these redshifts. However, on average, cluster galaxies have a smaller \sizeratio{} ratio than field galaxies: $\langle$\sizeratio{}$\rangle = 0.96\pm0.09$ compared to $1.22\pm0.08$ (smaller at a 98\% credibility level). These values are uncorrected for the wavelength difference between \Ha{} emission and $R_c$-band stellar light, but implementing such a correction only reinforces our results. We also show that whilst the cluster and field samples follow indistinguishable mass-metallicity (MZ) relations, the residuals around the MZ relation of cluster members correlate with cluster-centric distance; galaxies residing closer to the cluster centre tend to have enhanced metallicities (significant at the 2.6$\sigma$ level). Finally, in contrast to previous studies, we find no significant differences in electron number density between the cluster and field galaxies. We use simple chemical evolution models to conclude that the effects of disc strangulation and ram-pressure stripping can quantitatively explain our observations.
~~~~~~~~~~~
\end{abstract}

\begin{keywords}
galaxies: clusters: general -- galaxies: evolution --  galaxies: ISM
\end{keywords}

\section{Introduction}

It is well understood that the environment in which a galaxy resides plays an important role in its formation and evolution. Focusing on the densest environments in particular, we have known for many years that the galaxy population residing in galaxy clusters is markedly different from its counterpart in the field: galaxy cluster members tend to have early-type morphologies \citep{Dressler:1980, Dressler:1997}, redder optical colours \citep[e.g.][]{Pimbblet:2002} and spectra free of emission lines \citep{Gisler:1978}. Current work has extended these observations to much higher redshifts, with studies of protoclusters and overdensities at redshifts between $1.5<z<2.5$ becoming common (e.g. \citealt{Muzzin:2013}; \citealt{Shimakawa:2015}; \citealt{WangT:2016}; \citealt{Prichard:2017}; \citealt{Perez-Martinez+2017}; \citealt{Boehm:2019}; see \citealt{Overzier:2016} for a review). 

The physical processes which cause the differences in galaxy properties can be broadly separated into two categories. On one hand, a number of "external" mechanisms acting on cluster galaxies (involving their interactions with the intracluster medium or other cluster members) have been suggested to quench their star formation and alter their properties. Of these, perhaps the most dramatic is ram-pressure stripping \citep[first proposed in][]{Gunn:1972}. Galaxy clusters are the largest potential wells in the Universe, and contain vast quantities of hot gas between their members (see e.g. \citealt{Sarazin:1986} and \citealt{Kravtsov:2012} for reviews). This intracluster medium (ICM) contains an order of magnitude more mass than is in the stars of the galaxies themselves, and is around a thousand times more dense than the inter-galactic medium which surrounds galaxies outside clusters \citep[e.g.][]{Nicastro:2008, Zhuravleva:2013}. When a galaxy falls into a cluster, its motion through the ICM creates a pressure which acts on its reservoirs of gas. The force exerted can be strong enough to overcome the disc's gravitational restoring force, stripping away this reservoir in an occasionally spectacular fashion. Direct observational evidence of gas being stripped from cluster galaxies can be found at local and intermediate redshifts \citep[e.g.][]{Owers:2012, Ebeling:2014,Rawle:2014, Poggianti:2017, Boselli:2019}, with such objects coming to be known colloquially as "Jellyfish" galaxies following \cite{Smith:2010}.  

On the other hand, galaxy clusters are inherently special places, and the initial conditions of galaxies that form within them are different from those of galaxies which form in less dense regions of space. Since the massive clusters of today correspond to the largest overdensities in the early Universe \citep[e.g.][]{Springel:2005}, it has been suggested that these unique initial conditions lead to an "accelerated" evolution of their members \citep[e.g.][]{Dressler:1980, Morishita:2017, Chan:2018}. The question of whether internal or external drivers of galaxy evolution are most important is key to building a complete picture of the way in which galaxies change throughout their lifetimes, and a satisfactory answer has so far remained out of reach.  

Attempting to answer this question by studying cluster galaxies at $z=0$ is hampered by the fact that so many of them are quiescent, evolved and seemingly at the endpoint of their evolutionary paths. As first discussed in \cite{Butcher:1978, Butcher:1984}, galaxy clusters at $z\approx0.5$ contain a much higher fraction of star-forming galaxies than today. Furthermore, of those cluster members which are not currently forming stars, some show evidence of recently truncated star formation via the k+a spectral characteristics of post-starburst galaxies \citep[e.g.][]{Poggianti:2009} or the strong H$\delta$ absorption of "post star-forming" galaxies \citep[e.g.][]{Couch:1987, Owers:2019}. These observations imply that intermediate-redshift clusters-- which are more likely to be in the process of actively transforming their members-- offer a more promising route to address this problem. 

A number of studies have targeted intermediate-redshift cluster galaxies, often with spatially-unresolved spectroscopy \citep[e.g. recently][]{Rosati:2014, Sobral:2015, Maier:2016, Morishita:2017}. Whilst these studies have the advantage of targeting large numbers of objects and forming statistically-significant sample sizes, environmental quenching processes are inherently spatially inhomogeneous. Spectroscopic observations which sample multiple positions in the same galaxy at the same time are therefore required to catch these mechanisms to transform galaxies in the act. 

Our view of intermediate- to high-redshift ($z>1$) star-forming galaxies has been revolutionised in the last decade by integral-field spectroscopic surveys from the ground \citep[e.g.][]{ForsterSchreiber:2006, Mancini:2011, Genzel:2011, Wisnioski:2015, Stott:2016, Beifiori:2017} and deep \textit{Hubble Space Telescope (HST)} grism spectroscopy \citep[e.g.][]{Atek:2010, vanDokkum:2011, Brammer:2012, Schmidt:2014, Treu:2015}. These surveys generally study ionised-gas emission, primarily the \Ha{} or [\ion{O}{iii}]$\lambda5007$ lines, resulting in spatially resolved maps of the rate and locations of star formation, two-dimensional maps of the gas kinematics and information on variations in the interstellar medium (ISM) conditions across individual objects. 

There are few studies of this kind, however, which specifically target star formation in cluster members with spatially-resolved spectroscopy.  Pioneering work in this field was carried out by \citet{Kutdemir:2008, Kutdemir:2010}, who used the Very Large Telescope (VLT) FOcal Reducer/low dispersion Spectrograph 2 (FORS2) with an observing pattern of three adjacent parallel slits to target ionised-gas emission in cluster and field galaxies at $0.1\leq z\leq0.91$. They found a remarkable similarity between the fraction of galaxies with irregular gas kinematics in their field and cluster samples, whilst also finding a correlation between \Ha{} luminosity and gas kinematic irregularity which only holds for cluster members. 

Two further recent examples are \cite{Vulcani:2015, Vulcani:2016}, who used data from the Grism Lens-Amplified Survey from Space (GLASS) to study the morphologies and star formation rates of 76 \Ha{} emitters in 10 clusters from the Cluster Lensing and Supernova survey with \textit{Hubble} \citep[CLASH; ][]{Postman:2012}. They found that \Ha{} emitters are observed with a wide variety of morphologies, including a number undergoing ram-pressure stripping, and that their cluster samples follow a mass--star formation rate (SFR) relation similar to that of a matched sample of galaxies in the field. 

In this work, we perform a similar investigation using observations from the ``K-CLASH" survey (\citealt{Tiley:2020}; hereafter \citetalias{Tiley:2020}). We target 4 clusters from the CLASH sample\footnote{Only one, MACS 2129, is also studied in \cite{Vulcani:2016}} using the $K$-band Multi-Object Spectrograph \citep[KMOS;][]{Sharples:2013}, building a sample of 40 star-forming cluster galaxies and 120 star-forming mass-matched "field" galaxies along the same lines-of-sight. Our goal is to find evidence of environmental quenching mechanisms in action, by comparing the properties of these star-forming cluster and field galaxies which have been observed in a homogeneous way. 

In Section \ref{sec:K-CLASH_survey}, we briefly summarise the K-CLASH survey (introduced fully in \citetalias{Tiley:2020}) and define the cluster and field galaxy samples used in this work. In Section \ref{sec:Halpha_size}, we discuss our method of creating and characterising our \Ha{} surface brightness distributions and the $R_\textrm{c}$-band continuum images. We then measure the half-light radii of the \Ha{} and $R_\mathrm{c}$-band images ($r_{\mathrm{e}, \rm{H}\alpha}$ and $r_{\mathrm{e}, R_{\mathrm{c} } }$  respectively) and present distributions of the ratio \sizeratio{} for the cluster and field galaxies. In Section \ref{sec:Stacked_spectra} we describe our method of measuring emission line fluxes and line ratios from individual galaxies, as well as our spectral stacking methodology, and discuss the results. We place our results into context and discuss the physical implications of our findings in Section \ref{sec:Discussion}, before drawing our conclusions in Section \ref{sec:Conclusions}. 

We use the program \texttt{Stan}\footnote{\url{https://mc-stan.org/}} \citep{Carpenter:2017} a number of times in this work to perform full Bayesian inference of model parameters via dynamic Hamiltonian Monte Carlo (HMC) sampling \citep{Betancourt:2014, Betancourt:2017}. In each case, we ensured that the Gelman-Rubin convergence statistic $\hat{R}$ was within acceptable ranges (i.e. below 1.1 for each parameter) and there were no divergent transitions during the sampling. All magnitudes referred to in this work are in the AB system. We assume a Wilkinson Microwave Anisotropy Probe (WMAP) nine-year cosmology \citep{WMAP9} with Hubble constant $\mathrm{H}_0$ = 69.3\kms Mpc$^{-1}$, matter density $\Omega_{\mathrm{m}} = 0.287$, spatial curvature density $\Omega_{k}=0$ and cosmological constant $\Omega_{\Lambda}$ = 0.713.

\section{The K-CLASH survey}
\label{sec:K-CLASH_survey}

The K-CLASH survey design, data reduction procedures and sample properties are introduced and described in \citetalias{Tiley:2020}. We provide a brief summary here. 

KMOS is a multi-object near-infrared spectrograph mounted at the Nasmyth focus of Unit Telescope 1 (UT1) at the European Southern Observatory's (ESO) VLT, Cerro Paranal, Chile. It consists of 24 separate integral-field units (IFUs) on pick-off arms which can be deployed anywhere within a 7\farcm2 diameter patrol field; each IFU itself has a field of view of 2\farcs8 $\times$ 2\farcs8 with spatial sampling of 0\farcs2 $\times$ 0\farcs2 per spaxel.

K-CLASH observations were conducted with KMOS in the \textit{IZ} band between 2016 and 2018 (proposal IDs 097.A-0397, 098.A-0224, 099.A-0207 and 0100.A-0296). The wavelength coverage in the \textit{IZ} band is from $0.779$ to 1.079 $\mu$m, corresponding to rest-frame \Ha{} emission from $z=0.19$ to $z=0.64$. The resolving power varies from $R=2700$ at the bluest wavelength to $R=3700$ at the reddest. The data were reduced with the publicly available \textsc{EsoRex} software \citep[the "ESO Recipe Execution Tool" ;][]{esorex} and the KMOS instrument pipeline. The pipeline propagates uncertainties in the standard manner, resulting in a "noise" cube for each galaxy to accompany its "data" cube.

The target fields were chosen to be the four massive galaxy clusters MACS 2129 ($z=0.589$), MACS 1311 ($z=0.494$), MACS 1931 ($z=0.352$) and MS 2137 ($z=0.313$). These clusters were selected from the full CLASH sample\footnote{\url{https://archive.stsci.edu/prepds/clash/}} to be observable from the VLT and to be at redshifts where \Ha{} emission from cluster members lies between atmospheric telluric absorption bands and strong night sky emission-line features. A summary of the properties of each cluster is presented in Table \ref{tab:clashclusters}. Each cluster was also required to have wide-field optical imaging in multiple bands\footnote{Optical imaging is generally from Suprime-Cam \citep{Miyazaki:2002}, but is supplemented by data from the ESO Wide Field Imager \citep{Baade:1999} and the Magellan Inamori Magellan Areal Camera (IMACS) in MACS 1311 where only $R_\mathrm{c}$-band Suprime-Cam imaging was available.}, from which we select bright galaxies (\textit{V} < 22 for MACS 1931 and \textit{V} < 23 otherwise)  with good photometric redshift estimates (measured by \citealt{Umetsu:2014}) as targets. We preferentially observed galaxies which are blue ($B - V\leq 0.9$ for $z \leq 0.4$ and $V - R_\mathrm{c} \leq 0.9$ for $z > 0.4$) and have photometric redshifts placing them at their respective cluster redshift.  Remaining KMOS arms were first placed on blue galaxies at other photometric redshifts, followed by red galaxies at the cluster redshift. During every observing block (OB), one KMOS IFU was allocated to the brightest cluster galaxy (BCG), and at least one arm was placed on a star in the field of view to measure the point-spread function (PSF). Whilst multi-band \textit{HST} photometry is available in each cluster centre, the limited radial extent of these observations means only a small fraction of our K-CLASH targets are covered, and as such we do not make use of this photometry in this work. 

In total, 282 galaxies were observed across the 4 clusters. We detected stellar continuum and/or ionised-gas emission (from the \Ha{} and/or [\ion{N}{ii}] lines) in 243 galaxies. As discussed in \citetalias{Tiley:2020}, after integrating each KMOS observation in 0\farcs6, 1\farcs2, and 2\farcs4 diameter apertures, we measured the emission-line signal-to-noise ratio (S/N) by simultaneously fitting the \Ha{} line and each of the [\ion{N}{ii}] doublet lines with a Gaussian component. Following \cite{Stott:2016}, \cite{Tiley:2016} and \cite{Tiley:2019}, the S/N of the \Ha{} emission is then defined as the square root of the difference in $\chi^2$ between that of the best-fitting \Ha{} Gaussian component ($\chi^2_{\mathrm{model}}$) and that of a straight line equal to the value of the continuum ($\chi^2_{\mathrm{
continuum}}$); i.e. S/N = $\sqrt{\chi^2_{\mathrm{model}} - \chi^2_{\mathrm{continuum}}}$. \Ha{} emission with S/N>5.0 in at least one aperture was found in 191 objects, forming the K-CLASH parent sample. 

\begin{table*}
\centering
\caption{Properties of the CLASH clusters observed as part of the K-CLASH survey.  Columns refer to: (1) the (abbreviated) cluster name from the CLASH survey; (2 and 3) the right ascension and declination of the brightest cluster galaxy (BCG) in each cluster; (4) the redshift of the BCG; (5) the cluster X-Ray temperature; (6) the radius at which the mean interior density of the cluster is 200 times the critical density of the Universe; (7) the number of galaxies targeted with KMOS; (8) the number of galaxies detected with \Ha{} S/N greater than five; (9) the number of galaxies from each field which are members of the K-CLASH cluster sample; and (10) the number of galaxies from each field which are members of the K-CLASH field sample. Superscripts refer to: (a) published values from \protect\cite{Postman:2012}; (b) radii derived from mass models initially constructed by \protect\cite{Zitrin:2009, Zitrin:2013} but later updated (A. Zitrin, private communication).}
\begin{tabular}{ llcccccccc}
\hline
Cluste & RA& Dec & $z$\textsuperscript{(a)} & $kT_{X}$$^{\rm{(a)}}$ &   $R_{200}^{\rm{(b)}}$ & KMOS targets& \Ha{} detections & Cluster sample & Mass-matched  \\
 &  (J2000) &  (J2000) &  &  (keV) & (kpc) & & & &Field sample\\
 \textbf{(1)}&\textbf{(2)}&\textbf{(3)}&\textbf{(4)}&\textbf{(5)}&\textbf{(6)}&\textbf{(7)}&\textbf{(8)}&\textbf{(9)}&\textbf{(10)}\\
\hline
MACS2129 & 21:29:26.12 & $-$07:41:27.76 & 0.589 & $9.00\pm1.20$  & 1904 & 75 & 57 & 12 & 39\\
MACS1311 & 13:11:01.80 & $-$03:10:39.68  & 0.494 & $5.90\pm0.40$ & 1395 & 76 & 44 & 15 & 22\\
MACS1931 & 19:31:49.63 & $-$26:34:32.51 & 0.352 & $6.70\pm0.40$ & 1871 & 63 & 44 & 4 & 30\\
MS2137 & 21:40:15.17 & $-$23:39:40.33 & 0.313 & $5.90\pm0.30$  & 1261 & 68 & 46 & 9 & 29\\
\hline
\end{tabular}
\label{tab:clashclusters}
\end{table*}

\subsection{Removal of AGN}
\label{sec:AGN_removal}

Since this work concentrates on star-forming galaxies, it is important to distinguish between ionised-gas emission which traces recent star formation and ionising photons from active galactic nuclei (AGN). Unfortunately, we cannot place our objects on many of the common emission-line diagnostic diagrams used to identify AGN contamination \citep[e.g. the BPT diagram:][]{BPT, Kauffmann:2003, Kewley:2006} due to the fact that the KMOS wavelength range does not encompass the H$\beta$ and [\ion{O}{iii}] emission lines for all of our targets. 

Instead, we turn to the [\ion{N}{ii}]/\Ha{} ratio as well as a number of sources of ancillary data: the Chandra Advanced CCD Imaging Spectrometer (ACIS) survey of X-ray Point Sources \citep{Wang:2016}, the Wide-field Infrared Survey Explorer \citep[WISE;][]{WISE, NeoWISE} "AllWISE" source catalogue\footnote{\url{http://wise2.ipac.caltech.edu/docs/release/allwise/}}, and \textit{Spitzer Space Telescope} Infrared Array Camera \citep[IRAC;][]{IRAC} observations in the 3.6 and 4.5$\mu$m channels\footnote{\url{https://irsa.ipac.caltech.edu/data/SPITZER/CLASH/}}. 

Using these sources, we remove from the parent K-CLASH sample: 

\begin{itemize}
\item 6 galaxies with X-ray luminosities between $10^{42}$ and $10^{44}$ erg s$^{-1}$, likely to be powered by an AGN \citep[e.g.][]{Comastri:2004}. Of these, 5 are detected in \Ha{} with S/N>5.
\item 13 galaxies with WISE colour $W_1-W_2>0.8$ (following the AGN selection criterion of \citealt{Stern:2012}). Of these, 10 have \Ha{} with S/N>5.
\item 1 galaxy with \textit{Spitzer} colour [3.6]-[4.5]>1.0, which is not detected in \Ha{}. Unfortunately, we were unable to use the common colour-colour cuts from \cite{Donley:2012}, since neither [5.8] nor [8.0] micron observations of our fields were available. 
\item 13 galaxies with emission line ratios $\log_{10}($[\ion{N}{ii}]/\Ha{})$>-0.1$ (following \citealt{Wisnioski:2018}). Note that these emission line fluxes are measured in a 1\farcs2 diameter circular aperture. 
\end{itemize}

Five galaxies with \Ha{} S/N>5 were classified as containing an AGN using two or more diagnostics. 

\subsection{Cluster and field samples}
\label{sec:sample_selection}

Next, we differentiate between galaxies that reside in one of the targeted CLASH clusters and those which are simply chance alignments along the same line of sight. 

Firstly, we calculated the predicted velocity dispersion ($\sigma_{\rm{cluster}}$) of each of the four CLASH clusters using the dispersion-temperature ($\sigma-T$) relation of \citet{Girardi:1996} and the cluster X-ray temperature from \cite{Postman:2012}. We verified that using the velocity dispersion predicted assuming a hydrostatic isothermal model ($\sigma^2=k_{\rm{B}}T/\mu m_p$\footnote{where $k_{\rm{B}}$ is the Boltzmann constant, $T$ is the gas temperature $m_p$ is the mass of a proton and $\mu=0.6715$ is the mean molecular weight in atomic mass units.}) or the $\sigma$-T relation of \cite{Wu:1999} made no difference to our sample selection. The cluster redshift ($z_{\rm{cluster}}$) was taken from \cite{Postman:2012} . For each galaxy with detected \Ha{} emission, we then used its spectroscopic redshift ($z_{\rm{member}}$) to calculate its line-of-sight velocity with respect to the rest-frame of the cluster as $v_{\rm{member}}=c(z_{\rm{member}}-z_{\rm{cluster}})/(1+z_{\rm{cluster}})$, where $c$ is the speed of light. 

Star forming galaxies (SFGs) with a projected radius ($r$) less than twice the radius where the mean interior density is 200 times the critical density of the Universe ($R_{\rm{200}}$) and with $|v_{\rm{member}}|$ less than three times $\sigma_{\rm{cluster}}$ are then classified as cluster members and form the K-CLASH cluster sample. We note that we use updated $R_{\rm{200}}$ measurements of the four CLASH clusters from A. Zitrin (private communication); these values are listed in Table \ref{tab:clashclusters}. The widths of these windows were a compromise to account for the possibility that the clusters may not be completely relaxed whilst minimising contamination from non-cluster members. We then selected the following populations:

\begin{itemize}
\item Galaxies which do not contain an AGN (based on the criteria of Section \ref{sec:AGN_removal}), have \Ha{} S/N>5 and satisfy $r < 2 R_{\rm{200}}$ and $|v_{\rm{member}}| < 3 \sigma_{\rm{cluster}}$ form the K-CLASH cluster sample. This selects 40 galaxies. 
\item Galaxies which do not contain an AGN, have \Ha{} S/N>5, do not satisfy both $r < 2 R_{\rm{200}}$ and $|v_{\rm{member}}| < 3 \sigma_{\rm{cluster}}$ and have $9.5 < \log_{10}(M_*/\mathrm{M}_{\odot}) < 11.1$ form the K-CLASH field sample. This selects 120 galaxies. 
\end{itemize}

A further 8 galaxies have $\log_{10}(M_*/\mathrm{M}_{\odot}) > 11.1$ and form a "high-mass field" sample. Due to their small number, however, we refrain from analysing them further in this work\footnote{Note that the "field" sample discussed in \citetalias{Tiley:2020} encompasses all 128 galaxies.}.  A detailed discussion of the properties of these samples can be found in \citetalias{Tiley:2020}. We note that these criteria are not perfect, and the fact that we have treated each cluster as axisymmetric is unlikely to be strictly correct. It is therefore possible that the field sample contains galaxies which actually reside in the cluster, and vice-versa. This implies that the differences between cluster and field galaxies found in this work are formally only lower limits, as any contamination at all (in either direction) will tend to homogenise both samples and wash out any differences we measure.

\section{Measuring H$\alpha$ and stellar continuum sizes}
\label{sec:Halpha_size}

Utilising the spatially-resolved nature of the KMOS observations of each galaxy, we now measure the extent of the H$\alpha$ emission in each object, a proxy for the spatial extent of star formation. In short, this involves measuring the point-spread function (PSF) of the observations, creating an H$\alpha$ emission line map and then fitting this map with a model light profile which has been convolved with the measured PSF. 

\subsection{The KMOS PSF}
\label{sec:KMOS_PSF}
The KMOS PSF represents the response of the instrument to a point-like input signal. Our observing strategy required at least one KMOS arm to target a star in the field of view for each OB, allowing us to measure the PSF. These stellar observations were then reduced in the same way as the science data (see \citetalias{Tiley:2020}), including co-adding multiple observations of the same object over separate OBs. We then collapse each reduced cube by summing along the wavelength direction to create an image, which we use during the fitting process (see Section \ref{sec:Ha_surface_brightness_profiles}).

To characterise variations of the PSF between nights, we fit a two-dimensional Gaussian model to each collapsed PSF image. We found the average FWHM of these images across all K-CLASH fields to be 0\farcs78, with a standard deviation of 0\farcs15. In Appendix \ref{sec:KMOS_PSF_variation}, we also investigate changes of the PSF as a function of wavelength and between each of the three KMOS spectrographs, finding the impact of these effects to be minimal for our study. 

\subsection{H$\alpha$ line-maps}
\label{sec:Ha_line_maps}

\begin{figure}
\begin{center}
\includegraphics[width=0.5\textwidth]{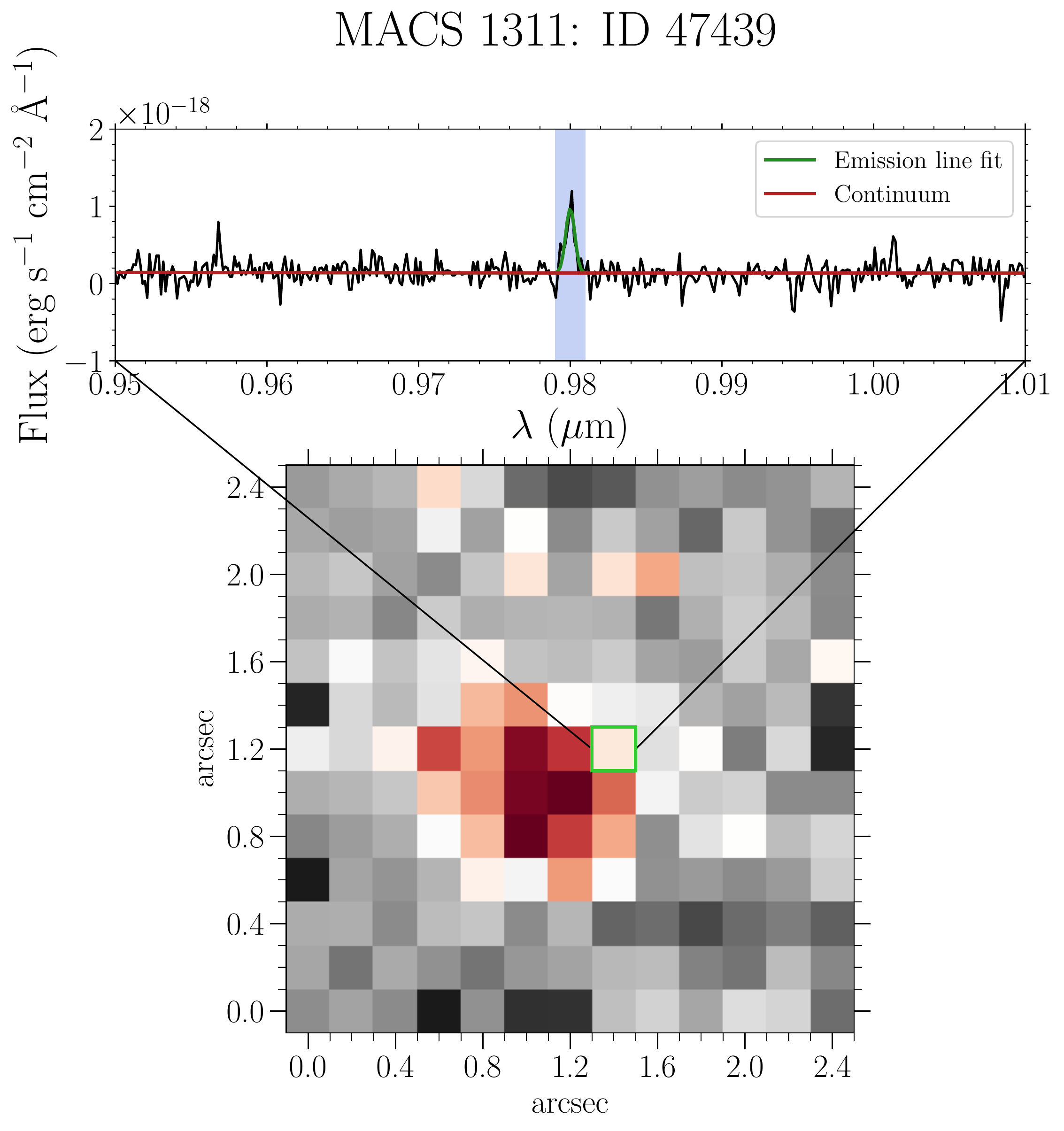}
\caption[An example \Ha{} line map]{An example \Ha{} line map for galaxy MACS 1311: ID 47439. The top panel shows a spectrum corresponding to a single spaxel (highlighted in green in the lower figure) with spectral range chosen to show the region around \Ha{} (blue), the emission line fit (green) and fit to the continuum (red).}
\label{fig:spaxel_example}
\end{center}
\end{figure}

To construct an \Ha{} line-map, we fit a Gaussian emission line profile to the spectrum in each spaxel after subtracting a 4\textsuperscript{th} order polynomial fit to the stellar continuum. We then integrate the best-fitting Gaussian to obtain the \Ha{} flux, and assign this value to the corresponding pixel of the \Ha{} image. To avoid including flux from the \ion{N}{ii} lines on either side, we mask 5 \AA{} regions around 6549.86 and 6585.27 \AA{} (their rest-frame wavelengths) during the fit. To avoid bad pixels, skylines and the \Ha{} emission itself biasing the continuum estimate, we iteratively sigma-clip the spectrum during the fitting, discarding pixels with discrepant fluxes and then fitting again to the remaining pixels. We again use a fourth order polynomial, fit with three iterations whilst discarding $\geq2\sigma$ outliers, but reasonable changes to these parameters do not affect our conclusions. We also create a corresponding two-dimensional "noise" image for each galaxy, using the galaxy's uncertainty cube. For each pixel in the noise image, we add in quadrature the values from the corresponding noise spectrum in a 20 \AA{} window around the \Ha{} line. This also allows us to make two dimensional S/N maps for each galaxy, by dividing its \Ha{} image its noise image\footnote{Note that this S/N definition is different to the one described in Section \ref{sec:K-CLASH_survey}  to measure the \Ha{} S/N in an integrated spectrum. }. An example \Ha{} line map is shown in Figure \ref{fig:spaxel_example}, with a representative spectrum showing the continuum and \Ha{} spectral regions.

An important consideration when measuring the size of \Ha{} emission is the minimum surface brightness our observations are sensitive to. We estimate the limit to which we can detect \Ha{} emission in each galaxy by taking the median of each wavelength slice in the error cube (thus creating a "median noise spectrum" for each target) and integrating this across a small window centred on the expected wavelength of the \Ha{} emission. The average FWHM of \Ha{} emission across the K-CLASH star-forming sample is 8.5 \AA, so we choose a window of 10 \AA; this window size also avoids contribution from [\ion{N}{II}] emission on either side. Our 3$\sigma$ detection limit is on the order of $1\times10^{-15}$ erg s$^{-1}$ cm$^{-2}$ arcsec$^{-2}$, which varies from $4\times10^{-15}$ to $4\times10^{-16}$ erg s$^{-1}$ cm$^{-2}$ arcsec$^{-2}$. To put this in context, this value is around three orders of magnitude shallower than the recent Multi-Unit Spectroscopic Explorer (MUSE) study of ram-pressure stripping at $z\approx0.7$ by \cite{Boselli:2019}, or the deep stacked \Ha{} images from \textit{HST} grism spectroscopy at $z\approx1$ studied in \cite{Nelson:2016}, which both reached a surface brightness limit on the order of $1\times10^{-18}$ erg s$^{-1}$ cm$^{-2}$ arcsec$^{-2}$ (although we note that the $\approx$100 \AA{} spectral resolution of the grism spectra studied in \citealt{Nelson:2016} is much lower than that of our KMOS spectra). 

\begin{figure}
\begin{center}
\includegraphics[width=0.5\textwidth]{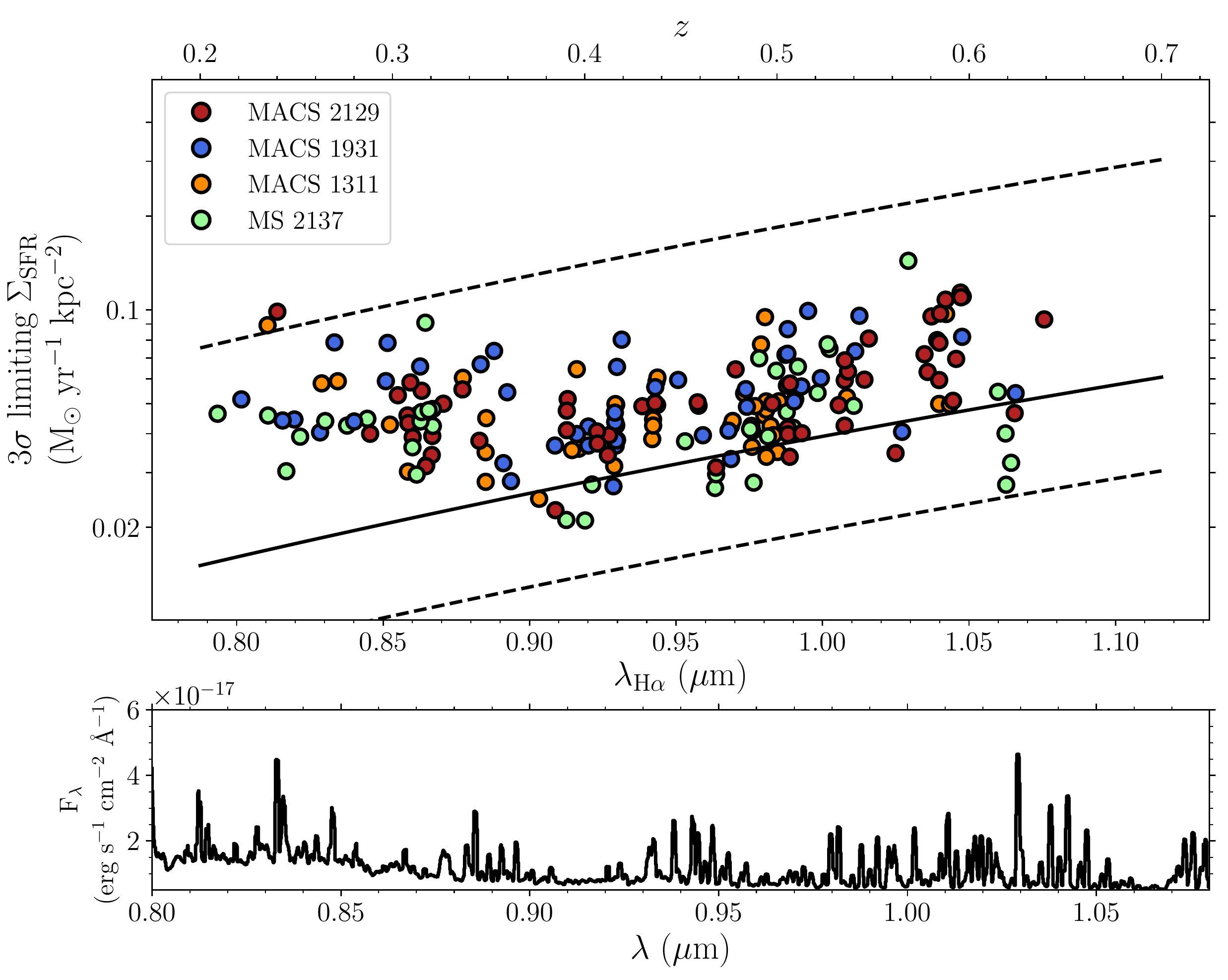}
\caption[Limiting surface brightness]{Top: The limiting star formation rate surface density ($\Sigma_{\rm{SFR}}$) as a function of the observed \Ha{} wavelength of each galaxy in our sample ($\lambda_{\mathrm{H}_{\alpha}}$). Data-points are colour-coded according to the CLASH field the galaxy was observed in. Solid and dashed lines show how a fixed background surface brightness (of $4\times10^{-15}$,  $7.5\times10^{-16}$ and $4\times10^{-16}$ erg s$^{-1}$ cm$^{-2}$ arcsec$^{-2}$) translates to a variable $\Sigma_{\rm{SFR}}$ (in M$_{\odot}$ yr$^{-1}$ kpc$^{-2}$). Note that this plot includes all \Ha{} detections, including those with S/N $<5$. Bottom: a representative spectrum showing how the noise level varies as a function of wavelength (due to strong sky emission lines and decreased sensitivity at the blue end of the \textit{IZ} band). To create this plot, we take a representative noise spectrum and find the average noise value (in a rolling 10 \AA{}  window) as a function of wavelength. The observations with lowest background noise (around 0.92 and 1.065 $\mu$m) correspond to gaps between skylines. Low-$z$ observations are affected by the higher average noise level at wavelengths $< 0.9\mu$m due to the decreased sensitivity of the KMOS detectors in the \textit{IZ} band. }
\label{fig:limiting_SFR_surface_density}
\end{center}
\end{figure}

We also convert these minimal \Ha{} surface brightnesses to minimal SFR surface densities ($\Sigma_{\rm{SFR}}$) using the relation of \cite{Hao:2011} and \cite{Murphy:2011} without correcting for dust extinction. The top panel of Figure \ref{fig:limiting_SFR_surface_density} shows the $3\sigma$ limiting $\Sigma_{\rm{SFR}}$ of each galaxy in our sample, which is $\approx$0.03-0.1 M$_{\odot}$ yr$^{-1}$ kpc$^{-2}$ for galaxies with an observed \Ha{} wavelength $\lambda_{\rm{H}\alpha}>0.9 \mu$m. The bottom panel shows how a representative noise spectrum (averaged spectrally in a rolling window of width 10 \AA{}) varies with wavelength. The decreased sensitivity of the detectors in the KMOS \textit{IZ} band (corresponding to a higher mean noise level shown in the bottom panel of Figure \ref{fig:limiting_SFR_surface_density}) implies that our observations of galaxies at $z\approx0.3$ are roughly as shallow in terms of SFR surface density as those at $z\approx0.5$. Furthermore, our best observations occur at redshifts corresponding to gaps between prominent sky emission lines at $\lambda\approx0.92$ \AA{} and $\lambda\approx1.065$ \AA. Our $\Sigma_{\rm{SFR}}$ sensitivities are comparable to those of other IFU studies at $z\sim1-2$ \citep[e.g.][]{Genzel:2011, Stott:2016}, although shallower than \textit{HST} grism studies at intermediate redshifts (e.g. \citealt{Vulcani:2015} reach 0.01 M$_{\odot}$ yr$^{-1}$ kpc$^{-2}$).

\subsection{\Ha{} surface brightness profiles}
\label{sec:Ha_surface_brightness_profiles}

\begin{figure*}
\begin{center}
\includegraphics[width=\textwidth]{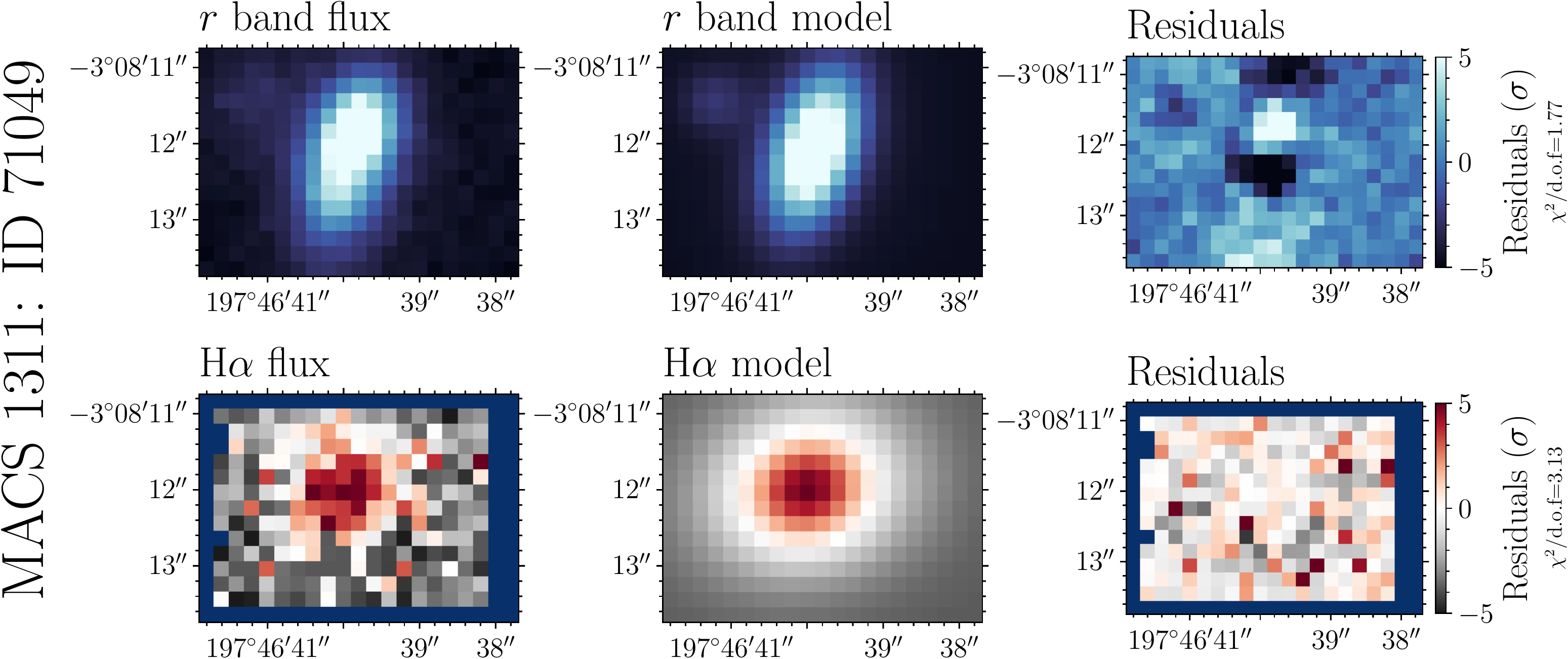}
\\
~
\\
\includegraphics[width=\textwidth]{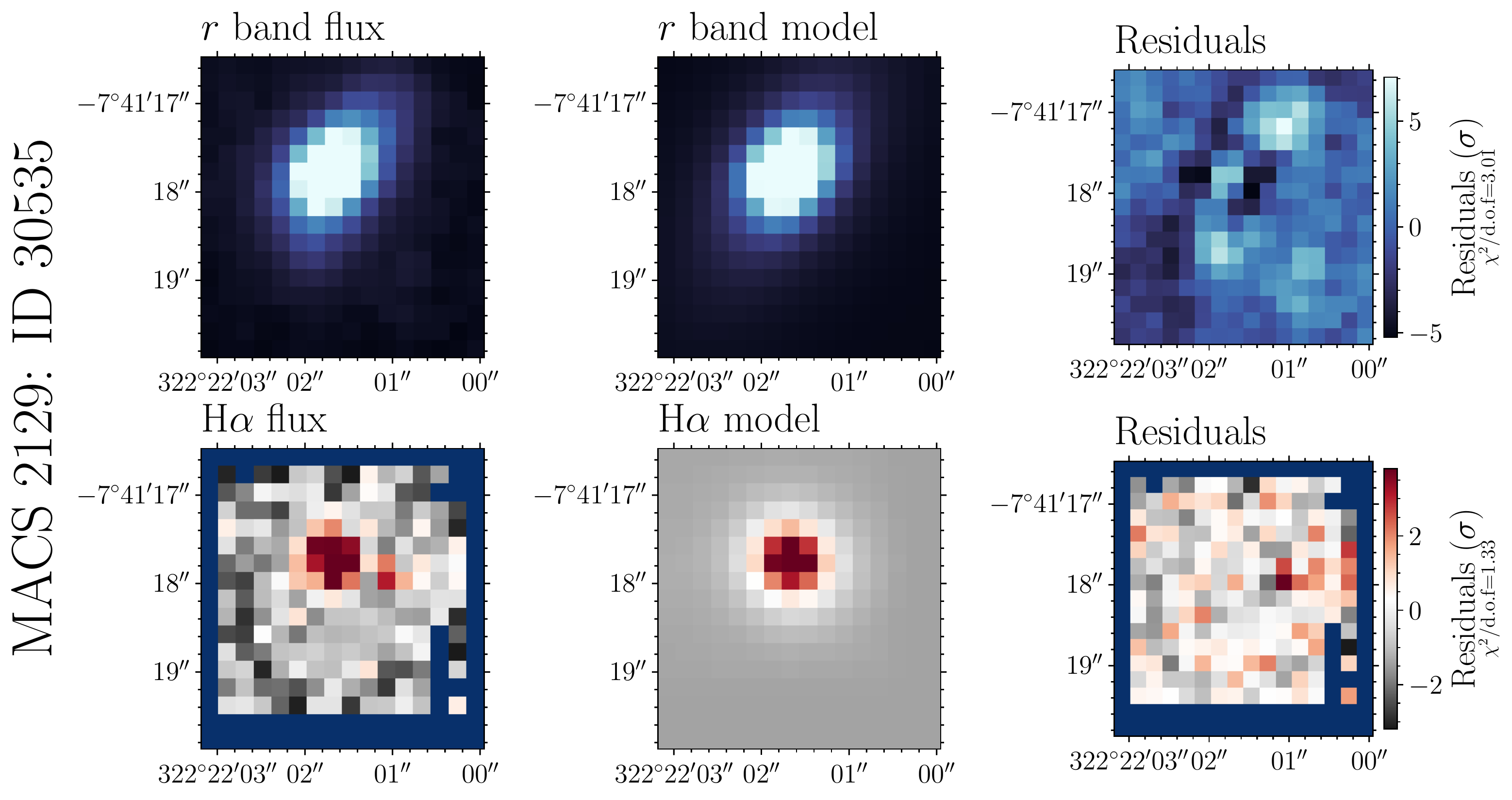}

\caption[]{Postage-stamp images extracted from the $R_\mathrm{c}$-band imaging (blue colour scales) and \Ha{} emission maps (red colour scales) for two galaxies from our sample. The first column shows the data, the second shows the model and the third shows the residuals from the fit. Note that the $R_\mathrm{c}$-band images have been shifted to match the centres of the \Ha{} emission maps.  The top object has an \sizeratio{} markedly greater than unity (extended \Ha{} emission compared to the stars), whilst the bottom panel shows an object with centrally concentrated \Ha{} emission.}

\label{fig:big_small_Ha}
\end{center}
\end{figure*}

With high-resolution broad-band and narrow-band imaging, the spatial structure of the continuum light \citep[e.g.][]{Elmegreen:2005} and \Ha{} emission \citep[e.g.][]{Shapley:2011, Nelson:2012} in some high-redshift galaxies has been found to be clumpy and disturbed, in contrast to the generally ordered stellar distributions and star formation seen throughout spiral galaxies today. For this reason, a number of studies use a curve-of-growth method to estimate the half-light radius of a galaxy's \Ha{} flux \citep[e.g.][]{Nelson:2012, Magdis:2016, Schaefer:2017}. This approach has some drawbacks, however. Low signal-to-noise \Ha{} flux at the outskirts of the galaxy may be missed, and spurious "hot" pixels in the line map are counted as real \Ha{} emission if not properly masked. Furthermore, the seeing-limited nature of our observations implies that we are also unable to distinguish distinct clumps of \Ha{} emission on scales smaller than the PSF, making irregular systems appear to follow smooth, disc-like surface brightness profiles. We therefore choose instead to fit model light profiles to our \Ha{} line maps, in a similar manner to \cite{Nelson:2016} and \cite{Wisnioski:2018}. We also note that this choice will have less of an impact on our intermediate-redshift study ($0.2\lessapprox z \lessapprox0.6$) than on work at higher redshift where disturbed morphologies are more common. 

We fit the \Ha{} surface brightness distributions using the publicly available code \imfit{}\footnote{\url{http://www.mpe.mpg.de/~erwin/code/imfit/}} \citep{Erwin:2015}. \imfit{} creates two-dimensional surface brightness distributions and fits these to data through a choice of minimisation techniques. In this case, we use \imfit{} to fit infinitely-thin axisymmetric exponential-disc surface brightness distributions to our \Ha{} line maps. Each model is convolved with an observation of the PSF (a two-dimensional image of a bright star in the field, constructed by collapsing the full KMOS data-cube along the wavelength direction; see Section \ref{sec:KMOS_PSF}) during the fitting process. Uncertainties on the fitting parameters are estimated using 1000 bootstrap resamples of the original image. We test the robustness of this fitting method in Appendix \ref{sec:SN_tests} and show it can accurately recover the input disc scale lengths from mock data at various values of signal-to-noise ratio, disc size and disc orientation. The intrinsic (i.e. deconvolved) \Ha{} half-light radius of each galaxy was then measured by performing a curve-of-growth analysis in circular apertures on the intrinsic (i.e. before convolution) best-fitting surface brightness profile. 

Each KMOS IFU has a field of view of 2\farcs8{}$\times$2\farcs8, which corresponds to 9.5$\times$9.5 and 20.3$\times$20.3 kpc at the lowest and highest redshift of our targets, respectively. We note that, in their study of \Ha{} emitters at $0.3<z<0.7$, \cite{Vulcani:2016} found no objects with \Ha{} effective radii larger than 10 kpc, implying that a KMOS IFU would encompass at least the half-light radius for all of their targets if observed at the highest redshift of our study. At the lowest redshifts, it is possible that some galaxies would have \Ha{} emission more extended than the field of view of an IFU.  As we show in Appendix \ref{sec:SN_tests}, however, we are able to recover the sizes of mock \Ha{} distributions larger than the field of view from high S/N data. We therefore conclude that whilst it is possible the field of view of a KMOS IFU may miss flux from the most spatially extended \Ha{} emitters, this is unlikely to significantly affect our conclusions. 

\subsection{Continuum imaging}
\label{sec:continuum_imaging}

Each K-CLASH field has been targeted with deep Subaru Suprime-Cam observations. Suprime-Cam \citep{Miyazaki:2002} has a 34$^{\prime}\times27^{\prime}$ field of view which was mosaiced over each cluster. The data were reduced and analysed by \cite{Umetsu:2014}, as well as independently by \cite{vonderLinden:2014}, using reduction methods described in \cite{Nonino:2009} and \cite{Medezinski:2013}. The images are publicly available from the CLASH archive\footnote{\url{https://archive.stsci.edu/prepds/clash/}}.  \textit{HST} imaging in a large number of bands is also available in the very centre of each CLASH cluster, but since this imaging covers only a small fraction of K-CLASH galaxies we choose not to use it for any of our targets. 

We use the Suprime-Cam $R_\mathrm{c}$-band imaging (in the Johnson--Morgan--Cousins system; see \citealt{Miyazaki:2002} for details) to measure the surface brightness profile of each K-CLASH galaxy. Images taken in this band are available without having been convolved to the limiting PSF of the other bands (``PSF-matched") before stacking. Instead, the $R_\mathrm{c}$-band images we use were stacked individually at each epoch and camera rotation angle, making them most appropriate for measuring galaxy shapes (e.g. for a weak lensing analysis) and light profiles. At the redshifts of our targets, the $R_\mathrm{c}$ band corresponds to the rest-frame $B$ band. 

Similarly to the \Ha{} line-maps, we used \imfit{} to fit model profiles to each galaxy. A $6\farcs4\times6\farcs4$ postage-stamp image of each target was extracted from the larger Suprime Cam $R_{\mathrm{c}}$-band image for this purpose. We note that this cutout is larger than a KMOS IFU field of view of $2\farcs8\times2\farcs8$; we found the results of our $R_{\mathrm{c}}$-band fitting to be more robust with this larger cutout size than when matching the KMOS IFUs' fields of view exactly. 

A median stack of hundreds of stars in each field was used as the PSF estimate during the fitting process. The seeing varied from 0\farcs6 to 0\farcs9 in the four K-CLASH fields. In contrast to the \Ha{} spatial modelling, however, we allowed the S\'ersic index of the light profile to vary between 1 and 10. We also simultaneously fit foreground and background objects in each postage-stamp cutout with appropriate S\'ersic or PSF models to ensure acceptable fits. Uncertainties were again measured using 1000 bootstrap resamples of the input data. Almost all galaxies were well fit with a single S\'ersic component, but a small number of disturbed objects required multiple components in order to achieve an adequate fit. The intrinsic $R_{\mathrm{c}}$-band half-light radius of each galaxy was again measured by performing a curve-of-growth analysis on the intrinsic best-fitting model (i.e. the model unconvolved with the PSF) in circular apertures.

\subsection{Signal-to-noise constraints}
\label{sec:SN_constraints}
Each galaxy has, up to now, been selected from the K-CLASH parent sample to have an \Ha{} S/N greater than 5 in at least one of a 0\farcs6, 1\farcs2 or 2\farcs4 diameter aperture, as well as not being flagged as containing an AGN (see Section \ref{sec:sample_selection}). To ensure our measurements from the image fitting are reliable, we now enforce further constraints. Firstly, we visually inspect each map and remove 2 galaxies from the cluster sample (5\%) and 10 from the field sample (8\%) where the fit has failed and/or there are problems with the $R_\mathrm{c}$-band imaging (e.g. the galaxy is obscured by the diffraction spike from a bright star). We then require that the reduced $\chi^{2}$ values of the \Ha{} and continuum fits should be less than 5\footnote{To ensure this constraint does not bias our results, we also conduct the same analysis without making this cut; see Section \ref{sec:Ha_Results}}. This removes 12 galaxies from the cluster sample (30\%) and 41 from the field sample (34\%).  Finally,  as motivated by our tests in Appendix \ref{sec:SN_tests}, we place a constraint on the minimum S/N of the \Ha{} images we use for further analysis. We divide each \Ha{} image by its  associated noise map (described in Section \ref{sec:Ha_line_maps}) to create a two dimensional S/N map for each galaxy, requiring that the average \Ha{} S/N within the best fitting half-light radius is greater than 2. This removes 4 galaxies from the cluster sample (10\%) and 21 from the field sample (18\%).  We are then left with 48 field galaxies and 22 cluster galaxies.

\subsection{Emission line--continuum size ratios}
\label{sec:Ha_Results}

 \begin{figure}
\begin{center}
\includegraphics[width=0.5\textwidth]{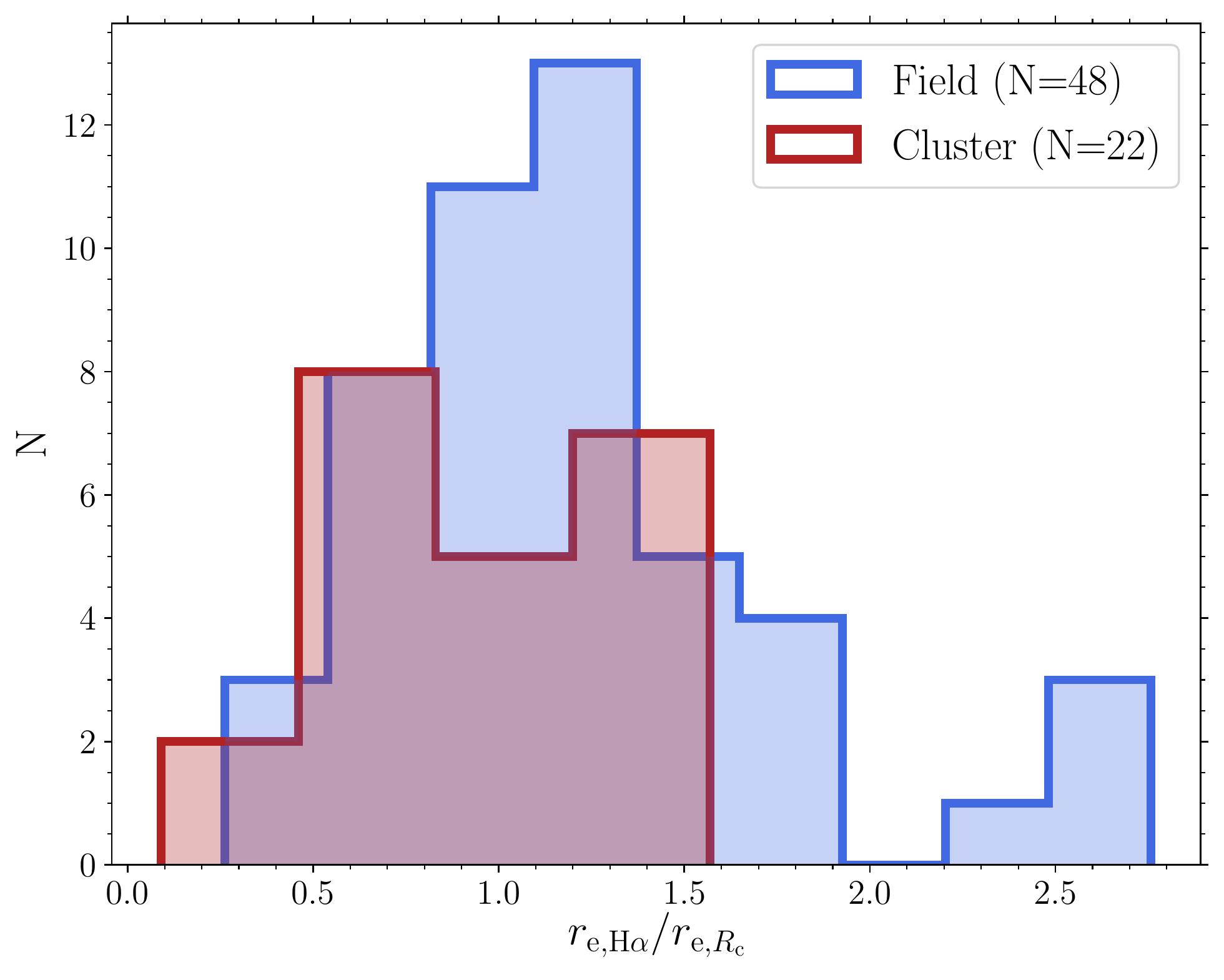}
\caption[]{Histograms of the ratio of \Ha{} size to continuum size for the cluster and field samples. The average \sizeratio{} ratio is $0.96\pm0.09$ for the cluster galaxies and $1.22\pm0.08$ for the field galaxies.}
\label{fig:size_ratio}
\end{center}
\end{figure}

Two example fits to the continuum and \Ha{} images are shown in Figure \ref{fig:big_small_Ha}. These objects were chosen to illustrate galaxies with extended (top) and concentrated (bottom) \Ha{} emission compared to their $R_\mathrm{c}$-band continuum size. The average ratio of \Ha{} effective radius to $R_\mathrm{c}$-band effective radius is 1.14$\pm$0.06, with a range from 0.1 and 2.76. This is in very good agreement with the study of \cite{Wilman:2020} at higher redshift, who found a median \Ha{}-to-continuum half-light radius ratio of 1.19 using the KMOS$^{3\mathrm{D}}$ sample \citep{Wisnioski:2015, Wisnioski:2019} at $0.7 < z < 2.7$. It is also in good agreements with the studies of $z\approx1$ galaxies by \cite{Nelson:2016} ($r_{\mathrm{s,\,H}\alpha}/r_\mathrm{s}[F140W] \approx 1.1$ for star-forming main sequence galaxies) and of "compact" SFGs by \cite{Wisnioski:2018} at $z\approx0.7-3.7$ ($r_{\mathrm{e,\,H}\alpha}/r_e[F160W] \approx 1.2$). Our measurements also agree well with the mass-(continuum) size relation at intermediate redshifts (\citealt{VanDerWel:2014}; see \citetalias{Tiley:2020}). 

 The average $r_{\rm{e, H}\alpha}/r_{\mathrm{e, } R_{\rm{c}}}$ ratio is $0.96\pm0.09$ for the cluster galaxies and $1.22\pm0.08$ for the field sample, where the quoted uncertainty is the standard error on the mean (i.e. the standard deviation of the sampling distribution). Figure \ref{fig:size_ratio} shows histograms of \sizeratio{} for the cluster and field samples. 
 
To ensure the constraints on $\chi^2$ introduced in Section \ref{sec:SN_constraints} are not biasing our results, we also compute the average \sizeratio{} ratio for the cluster and field sample without requiring a reduced-$\chi^2$ value of less than 5. In this case, our conclusions are unchanged; we find $\langle$\sizeratio{}$\rangle=1.03\pm0.09$ for the cluster galaxies (now including 30 galaxies) and  $\langle$\sizeratio{}$\rangle=1.55\pm0.20$ for the field sample (80 galaxies). 

We also test to see whether our choice of fitting each \Ha{} emission-line map with an exponential profile (equivalent to a S\'ersic profile with the index fixed at $n=1$) and each stellar continuum map with a S\'ersic profile (fitting the index as a free parameter) has impacted our results. We therefore repeat the above analysis using a S\'ersic profile for the \Ha{} maps, fixing the S\'ersic index of each galaxy to be the same as that measured for its continuum light. We find very similar results to before: the average \sizeratio{} ratio measured from this approach is $1.01+/-0.08$ for the cluster galaxies and  $1.26\pm0.09$ for the field galaxies.
  
The average \Ha{} half-light radii of the cluster and field galaxies are comparable: $\langle r_{e_{\rm{H}\alpha}}\rangle = 3.4\pm0.4$ kpc for the cluster galaixes compared to $3.9\pm0.3$ kpc for the field galaxies. The two samples also have consistent average $r_{e_{R_{c}}}$ values within the uncertainties: $3.5\pm0.3$ kpc for the cluster sample and $3.7\pm0.3$ kpc for the mass-matched field sample. 

It is well known that the measured size of an individual galaxy varies as a function of the wavelength it is observed at;  using data from the Galaxy And Mass Assembly (GAMA) survey \citep{Driver:2009, Driver:2011}, for example, \cite{Vulcani:2014} found that galaxies have smaller $r_{\mathrm{e}}$ when observed with redder photometric filters. Ideally, therefore, we would compare $r_{\mathrm{e}}$ from \Ha{} emission-line maps with $r_{\mathrm{e}}$ measured from deep stellar continuum imaging at $\approx 1\mu$m (which matches the \Ha{} rest-frame), rather than $r_{\mathrm{e}}$ from the $R_{\mathrm{c}}$-band images (rest-frame $B$ band) as done in this work. 

A number of studies have discussed empirical methods to convert size measurements carried out at one wavelength to size measurements at another, however \citep[e.g.][]{Kelvin:2012, VanDerWel:2014, Chan:2016}. We therefore use each of these prescriptions to correct our stellar-continuum effective radius measurements to the \Ha{} wavelength (6563 \AA) in each galaxy's rest-frame, and then recalculate the size ratios $r_{e, {\rm{H}\alpha}}/r_{e, R_{\mathrm{c}}\mathrm{\,(corrected)} }$ for the cluster and field galaxies.

The conversion from \cite{Chan:2016} applies identically to all galaxies. \cite{Kelvin:2012} and \cite{VanDerWel:2014} apply different corrections for disc-dominated/spheroid-dominated galaxies and late-type/early-type galaxies respectively. To use these prescriptions on our sample, we use the S\'ersic indices measured in Section \ref{sec:continuum_imaging} to classify galaxies as disc-dominated and late type (S\'ersic index $n \leq2$) or spheroid-dominated and early type (S\'ersic index $n > 2$).  We then recalculate the $r_{e, {\rm{H}\alpha}}/r_{e, R_{\mathrm{c}}\mathrm{\,(corrected)} }$ ratio for each galaxy. The average $r_{e, {\rm{H}\alpha}}/r_{e, R_{\mathrm{c}}\mathrm{\,(corrected)} }$ values are shown in Table \ref{tab:size_ratio_after_corrections}. For every correction prescription, the difference between the average size ratio of the cluster and field galaxies is as large or larger than the uncorrected difference, showing that our \sizeratio{} measurements are likely to be a lower limit on the true difference in size between \Ha{} emission and stellar continuum light at 6563\AA{} for cluster and field galaxies.

 \begin{table}
 \caption{Average $r_{e, {\rm{H}\alpha}}/r_{e, R_{\mathrm{c}}\mathrm{\,(corrected)} }$  ratios. Our measurements of stellar-continuum effective radii are from the rest-frame $B$ band of our targets, which is $\approx$ 2000\AA{} bluer than the rest-frame \Ha{} emission. Here, we apply a number of different prescriptions to correct our stellar effective radii to the rest-frame \Ha{} wavelength (6563\AA{}) and remeasure the size-ratio $r_{e, {\rm{H}\alpha}}/r_{e, \mathrm{continuum} }$. In each case, we find a difference between the average size ratio of the cluster and field galaxies which is as large or larger than that of the uncorrected case.}
\begin{tabular}{l | ccc}
Correction & Full sample & Cluster sample & Field sample \\
\hline
Uncorrected & $1.14\pm0.06$ & $0.96\pm0.09$ & $1.22\pm0.08$ \\
\cite{Chan:2016} & $1.28\pm0.07$ & $1.09\pm0.10$ & $1.37\pm0.09$ \\
\cite{Kelvin:2012} & $1.23\pm0.07$ & $1.04\pm0.10$ & $1.32\pm0.09$ \\
\cite{VanDerWel:2014} & $1.41\pm0.08$ & $1.19\pm0.12$ & $1.51\pm0.10$ \\
\end{tabular}
\label{tab:size_ratio_after_corrections}
\end{table}

\subsubsection{Statistical significance}

We have found that the mean \sizeratio{} of galaxies in our cluster sample is smaller than the average \sizeratio of galaxies in our field sample. Here, we quantify the significance of this result. 

We define $\Delta(\langle$\sizeratio$\rangle$) to be: 
$$
\Delta(\langle r_{\mathrm{e}, {\rm{H}\alpha}}/r_{\mathrm{e}, R_{\mathrm{c} } }\rangle) = \langle r_{\mathrm{e}, {\rm{H}\alpha}}/r_{\mathrm{e}, R_{\mathrm{c} } } \rangle_{\mathrm{cluster}} - \langle r_{\mathrm{e}, {\rm{H}\alpha}}/r_{\mathrm{e}, R_{\mathrm{c} } } \rangle_{\mathrm{field}} 
$$

\noindent and using the standard rules for addition or subtraction of two Gaussian random variables, our measurements in Section \ref{sec:Ha_Results} imply $\Delta(\langle r_{\mathrm{e}, {\rm{H}\alpha}}/r_{\mathrm{e}, R_{\mathrm{c} } }\rangle) = -0.26\pm0.12$. The 95\% credible interval for $\Delta(\langle r_{\mathrm{e}, {\rm{H}\alpha}}/r_{\mathrm{e}, R_{\mathrm{c} } }\rangle)$ therefore excludes zero (-0.49 $\leq\Delta(\langle$\sizeratio$\rangle)\leq$ -0.02), and using the cumulative distribution function of the normal distribution shows that only 1.5\% of the probability mass lies above zero. We therefore show that $\langle$\sizeratio$\rangle$ for the cluster galaxies is smaller than $\langle$\sizeratio$\rangle$ for the field galaxies at the 98.5\% credibility level. We have also verified this by direct simulation, as well as by fitting the cluster and field \sizeratio distributions with the program \texttt{Stan}\footnote{See the note in the introduction} and inspecting the posterior probability distribution of  $\Delta(\langle$\sizeratio$\rangle$). Finally, we perform a $t$-test (assuming unequal variances) with the null hypothesis that the cluster and field samples have equal means. The test shows we can reject this null hypothesis ($t \mathrm{\,statistic}=-2.12$, $p \mathrm{\,value}=0.038$).

\subsubsection{Comparison with previous work}
  
Our study is in very good agreement with the findings of \cite{Bamford:2007}. Their work measured the radial extent of rest frame $B$-band light and a number of emission lines ([\ion{O}{ii}]$\lambda3727$, H$\beta$, [\ion{O}{iii}]$\lambda4959$ and [\ion{O}{iii}]$\lambda 5007$)  in 19 cluster and 50 field galaxies at $0.25<z<1$. They found that the ratio of emission line to stellar scale length was 0.92$\pm$0.07 in cluster galaxies and 1.22$\pm$0.06 in field galaxies, in comparison to $0.96\pm0.09$ and $1.22\pm0.08$ from this sample. We also find agreement with \cite{Schaefer:2017}, who studied the half-light ratio $r_{50, \mathrm{H}\alpha}/r_{50, \mathrm{continuum}}$ in a low-redshift sample ($0.001 < z < 0.1$) of 201 star-forming galaxies in the Sydney-AAO Multi-object Integral-field (SAMI) survey. They found that at larger local environmental densities, the fraction of galaxies with centrally-concentrated \Ha{} emission (small \sizeratio{}) increases. 

Evidence for truncated \Ha{} discs in local galaxy clusters has also been reported by \cite{Koopmann:2006}, using narrow-band \Ha{} observations of spiral galaxies in the Virgo cluster. They found $r_{\rm{H}\alpha}/r_\mathrm{R}=0.91\pm0.05$, compared with $1.18\pm0.10$ for a matched sample of isolated spiral galaxies. This is again in excellent agreement with the averages found in this study, and indicates a lack of redshift evolution of the average ratio. \cite{Koopmann:2004}, and a follow up study by \cite{Crowl:2008}, also found that the star formation rates in the centres of truncated spirals in Virgo were comparable to a matched field sample, showing that these galaxies were undergoing stripping of their outskirts rather than experiencing a  reduction in star formation at all radii. 

On the other hand, \cite{Vulcani:2015,Vulcani:2016} found a small number of cluster galaxies at redshift $0.3<z<0.7$ with \textit{extended} \Ha{} compared to the stellar continuum, using \textit{HST} grism observations and rest-frame UV, optical and infrared \textit{HST} imaging. Of the galaxies with "spiral" morphologies in their sample, 7 out of 25 have \Ha{} sizes more than twice their size in the \textit{HST} $F475W$ band. They also concluded that these objects have been ram-pressure stripped, leading to an extended star-forming halo around the stellar component and complicated morphologies. As discussed in Section \ref{sec:Ha_line_maps}, the limiting SFR surface density of our observations is $\approx0.03-0.1$ M$_{\odot}$ yr$^{-1}$ kpc$^{-2}$, shallower than the studies of \cite{Vulcani:2015,Vulcani:2016} which reach 0.01 M$_{\odot}$ yr$^{-1}$ kpc$^{-2}$.  The reason for the apparent disagreement between the studies, therefore, could be due to the fact that we are insensitive to extended low surface brightness \Ha{} emission.

\subsection{What else could lead to small \sizeratio{} ratios? }

Previously, we have made the association of small \sizeratio{} ratios with the removal of \Ha{}-emitting gas at large galaxy radii. Here, we discuss other factors and processes which could lead to a similar trend whilst also explaining why their influence is expected to be small. 

The source of \Ha{} flux we want to measure is ultimately stars more massive than $\approx20$ \Msun{} and younger than 5-10 Myrs in individual star-forming regions (via their ionisation of surrounding gas; see e.g. \citealt{Kennicutt:1998a} and \citealt{Calzetti:2013} for reviews). We then want to make the association of \Ha{} flux and star formation (assuming a form for the initial mass function; see e.g \citealt{Kennicutt:2012}). Under the assumption that each \ion{H}{II} region is optically thick to ionising radiation (``Case B''--  every energetic photon from a massive star ionises an atom of Hydrogen, which then recombines and produces a cascade of emission lines, with \Ha{} and H$\beta$ the most prominent), a simple relation between \Ha{} flux and star formation rate exists \citep{Osterbrock:2006}. However, the effects of dust, AGN and photo-ionisation from evolved stars or shocks can confound this simple picture. 

Firstly, dust attenuation and extinction will suppress both \Ha{} and continuum flux, absorbing photons to be re-emitted at longer wavelengths. Is it possible that a difference in dust properties between cluster and field galaxies is driving the observed trend to small \sizeratio{} in dense environments? The $R_\mathrm{c}$ band in which we measure our continuum sizes is centred at $\approx6500$ \AA{}, tracing flux from 4000 to 5500 \AA{} in the rest-frame of our sources, 1000 - 2500 \AA{} bluer than rest frame \Ha{}. Dust reddening is a function of the emitted wavelength, with stronger extinction at shorter wavelengths \citep[e.g.][]{Calzetti:2000}, naively implying that dust will tend to attenuate the stellar continuum light more than the \Ha{} emission. On the other hand, however, a number of studies have reported additional attenuation towards star-forming regions \citep[e.g.][]{Fanelli:1988, Calzetti:1994, Mancini:2011}, finding that \Ha{} emission is further attenuated by a factor of $\approx2$ compared to the continuum at the same wavelength (A$_{V\mathrm{, H}\alpha}=$ A$_{V\rm{, continuum}}/0.44$). This is due to the fact that \ion{H}{II} regions, where young stars are found, are inherently dustier than the regions surrounding older stellar populations. 

Using the Calzetti extinction law \citep{Calzetti:2000} and taking these two effects into account, the \Ha{} emission is reddened more than the $R_\mathrm{c}$-band continuum light\footnote{For a representative extinction of A$_{\mathrm{V}}$=1 mag and a source at at $z=0.6$, \Ha{} is reddened by 1.87 mag compared to 1.36 mag for the continuum}. It is therefore possible that a difference in global dust properties between environments could lead to smaller observed size ratios in dense environments, if cluster galaxies are more obscured than their field counterparts. Using the global extinction estimates from our SED fits \citepalias{Tiley:2020}, however, we find the values of A$_{V}$ to be comparable between the (mass-matched) cluster and field samples, with the average being \textit{lower} for cluster galaxies by 0.1 mag. There is also no significant correlation between A$_{V}$ and size ratio (Pearson's correlation coefficient $r_{x,y}$=0.035, $p$-value=0.77), implying that galaxies with smaller size ratios are not systematically more attenuated. 

It should be noted that because the KMOS wavelength range available to us does not cover the H$\beta$ line, it is not possible to make \textit{local} extinction corrections to our \Ha{} maps (although differences in the spatial distribution of dust conspiring to suppress the \sizeratio{} ratio in cluster but not field galaxies is unlikely). We also note that whilst it is well known that dust extinction correlates with stellar mass \citep[e.g.][]{Reddy:2006,Momcheva:2013, Nelson:2016a}, we have avoided any associated systematic effect from this correlation by matching our field and cluster samples in mass.

Secondly, we consider the \Ha{} flux from active galactic nuclei. The effect of AGN would be to add extra flux in the centre of each galaxy, leading to centrally-peaked radial flux profiles and small inferred sizes. Narrow-line AGN in particular could impact the \Ha{} flux more than the continuum measurements, and hence bias the inferred  \sizeratio{} to small values. Whilst we have endeavoured to remove all AGN contamination from our sample (see Section \ref{sec:AGN_removal}), without further observations of the H$\beta$ and [\ion{O}{iii}] emission lines in each galaxy we cannot completely rule out their presence; in particular, weak AGN surrounded by star-forming regions are especially difficult to detect using the methods of Section \ref{sec:AGN_removal}. In the local Universe, the fraction of luminous AGN in high-density environments is lower than in the field \citep{Kauffmann:2004,Popesso:2006}, although there is strong evolution from $z>1$ to the present day \citep{Martini:2013}. For weaker AGN, the fraction tends to be comparable \citep{Best:2005a, Haggard:2010}. We therefore expect the effect of AGN interlopers which have been missed by our selection cuts to be small, but also-- most importantly-- comparable for the field and cluster sample.

Finally, photo-ionisation from sources such as planetary nebulae and post-AGB stars can contribute to \Ha{} emission \citep[e.g.][]{Binette:1994, Sarzi:2010}. Regions where such emission is an important fraction of the total ionising photon flux have come to be known as Low-Ionisation Emission Regions or "LIERs" \citep{Belfiore:2016}. The usual way to identify LIERs is with a BPT diagram, and as such we are unable to remove spectra with LIER-like line ratios from our samples. However, the total fraction of ionising radiation from post-AGB stars is largest for old, quiescent populations, so for the currently star-forming galaxies in our sample their contribution is expected to be small \citep{Byler:2017}.  

We therefore take our main result at face value: above the limiting surface brightness of our observations, star-forming galaxies residing in clusters on average have smaller \sizeratio{} values than similar galaxies in the field. 

\subsection{Central surface brightness measurements}

We now explore whether there are differences between the \Ha{} surface brightnesses in the cluster and field galaxies. 

We use the aperture flux measurements from \citetalias{Tiley:2020}, extracted from a 0\farcs6 diameter circular aperture centred on each galaxy. To briefly recap the measurement process, we first sum the spectra within the aperture to extract a spectrum and then subtract a sigma-clipped 6th order polynomial fit to the continuum after cleaning any remaining sky emission-line residuals. Fluxes are then measured by fitting a set of Gaussian emission line templates to the \Ha{} and [\ion{N}{ii}] emission lines (See Section 4.1 of \citetalias{Tiley:2020} for further details). Finally, we calculate the surface brightness of each galaxy within the 0\farcs6 aperture (which we denote $\mu_{0.6}$) by dividing the integrated flux of the \Ha{} line by the area of the aperture.

We note that using a fixed aperture size for all galaxies does not account for the fact that the apparent sizes of galaxies in the cluster and field samples are slightly different; the average \Ha{} half-light radius of galaxies in the cluster sample is 0\farcs56 whilst that of galaxies in the field sample is 0\farcs79. Whilst we would ideally measure the central surface brightness in apertures of e.g. $r_{\mathrm{e}}/4$ or $r_{\mathrm{e}}/8$, the average PSF width of the parent K-CLASH sample is comparable to$\approx$ 0\farcs6 and we therefore choose not to make measurements using apertures smaller than this. This does mean, of course, that measurements of $\mu_{0.6}$ for our cluster galaixes will include flux originating from slightly larger radii than for field sample galaxies (i.e. including flux from slightly beyond $r_{\mathrm{e}}$ on average for the cluster galaxies, compared to within $r_{\mathrm{e}}$ for the field galaxies). The central surface brightnesses $\mu_{0.6}$ for the cluster galaxies should therefore be taken as upper limits to the true (i.e. deconvolved) surface brightnesses within 0\farcs6.

\begin{figure}
\begin{center}
\includegraphics[width=0.5\textwidth]{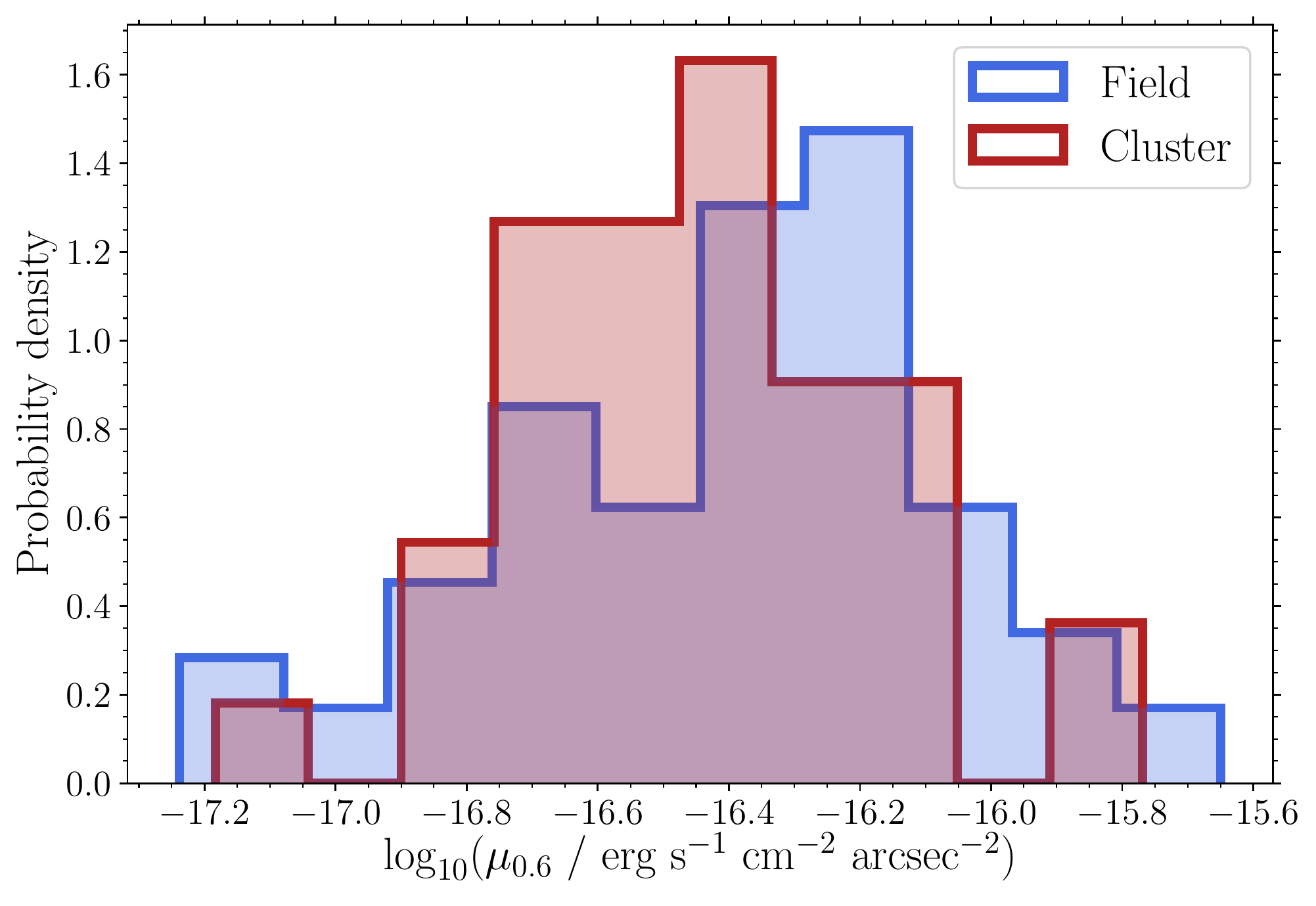}
\caption[]{Surface brightness measurements within a 0\farcs6 aperture ($\mu_{0.6}$) for the cluster (red) and field (blue) samples. We find a small difference between the average $\log_{10}(\mu_{0.6}/\mathrm{erg} \mathrm{\,s}^{-1}\mathrm{cm}^{-2}\mathrm{arcsec}^{-2})$ values:  $-16.29\pm{0.03}$ for the field sample and $-16.35\pm{0.05}$ for the cluster sample.}
\label{fig:surface_brightness}
\end{center}
\end{figure}

We find a small difference in the average central surface brightnesses ($\langle \mu_{0.6} \rangle$) between the two samples of $\approx$0.06 dex. We again use \texttt{Stan} to fit a Gaussian function to each distribution, measuring the average and standard deviation of each population (incorporating measurement uncertainties during the fit). The field sample is centred at $\log_{10}(\mu_{0.6}/\mathrm{erg} \mathrm{\,s}^{-1}\mathrm{cm}^{-2}\mathrm{arcsec}^{-2}) = -16.29\pm{0.03}$ and the cluster sample at          $\log_{10}(\mu_{0.6}/\mathrm{erg} \mathrm{\,s}^{-1}\mathrm{cm}^{-2}\mathrm{arcsec}^{-2}) = -16.35\pm{0.05}$. Histograms of the two distributions is shown in Figure \ref{fig:surface_brightness}. 

\section{Emission Line Analysis}
\label{sec:Stacked_spectra}

\begin{figure*}
\begin{center}
\includegraphics[width=\textwidth]{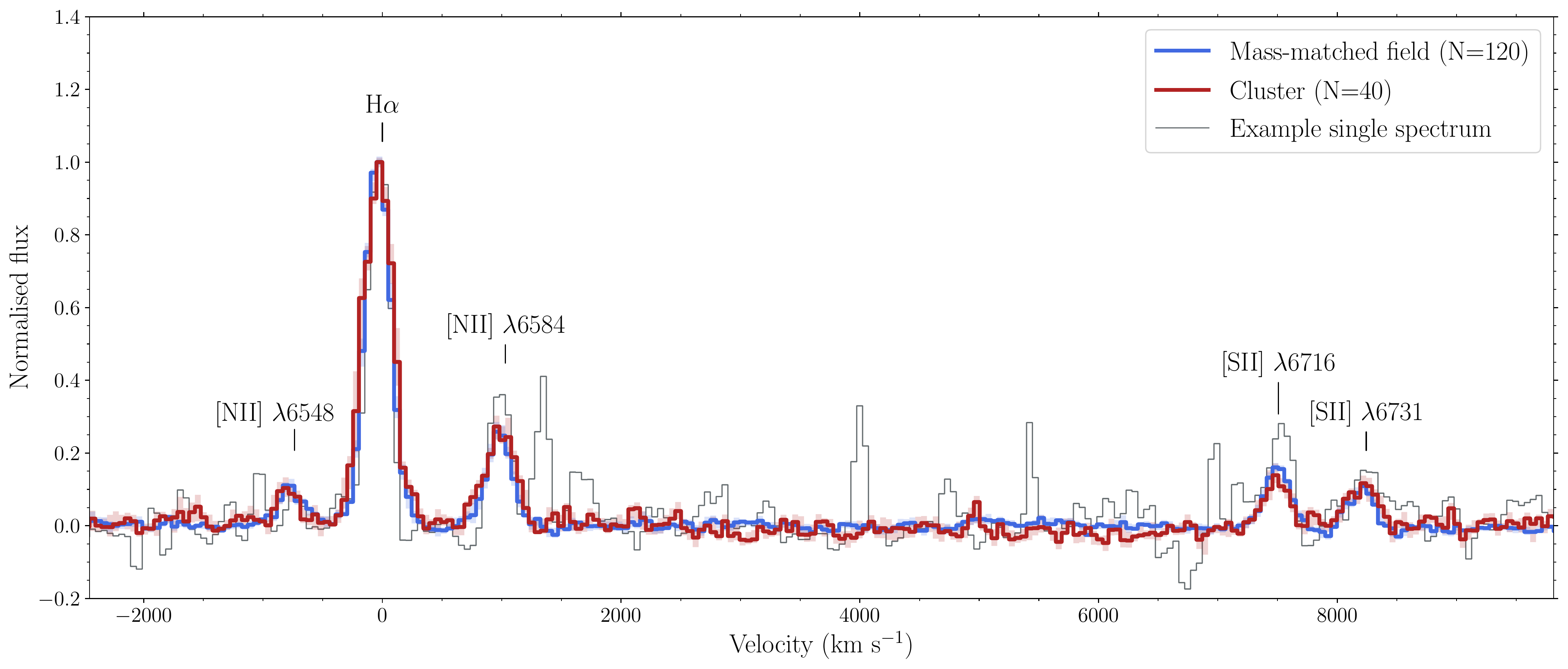}
\caption[Median stacked spectra for the field and cluster samples]{Median stacked spectra with normalised peak \Ha{} flux for the (mass-matched) field sample (blue) and cluster sample (red). Note that these spectra have not been convolved to a matching velocity dispersion ($\sigma$); $\sigma_{\rm{cluster}}=113$\kms{} and $\sigma_{\rm{field}}=98$\kms{}. We also show a representative spectrum from a single galaxy (grey). }
\label{fig:stacked_spectra}
\end{center}
\end{figure*}

Having found a difference in the \textit{extent} of star formation between cluster and field galaxies, we now assess whether the \textit{physical conditions} in their star-forming regions differ too. To do this, we investigate similarities and differences between the emission line spectra of galaxies in the cluster and mass-matched field sample.

We extract flux from a circular aperture with a diameter of 2\farcs4 centred on the continuum centre of each object. We measure the fluxes of the \Ha{}, [\ion{N}{ii}]$\lambda6548,\lambda6584$ and [\ion{S}{ii}]$\lambda6716,\lambda6731$ emission lines. This is accomplished by performing a fit (with a single velocity component) to each spectrum using \texttt{pPXF}\footnote{\url{https://pypi.org/project/ppxf/}} \citep{ppxf,2017MNRAS.466..798C}. The [\ion{N}{ii}] doublet is fit with a single template of two Gaussians, fixed at a flux ratio of 3 \citep{Osterbrock:2006}. The [\ion{S}{ii}] lines are fit with individual templates, but we use the "\texttt{limit\_doublets}" keyword in \texttt{pPXF} to limit the flux ratio of the two lines to be between 0.44 and 1.44, the values allowed by a physical analysis of the atomic transitions involved \citep{Osterbrock:2006}. We fail to detect stellar absorption features in our spectra, and as such use a sixth order polynomial to approximate the stellar continuum rather than including a library of stellar templates. We estimate uncertainties by adding random noise (scaled according to each galaxy's noise spectrum at each wavelength) to the best-fitting model and repeating the fit 1000 times per galaxy.

We also investigate a stack of the galaxy spectra in each of the cluster and field samples. Stacking increases the S/N compared to the spectra of individual galaxies, and allows us to make more robust measurements of the relatively faint [\ion{S}{ii}] doublet. 

During the stacking procedure, we interpolate each spectrum to be uniformly sampled in $\log\lambda$,  fit a Gaussian to find the centroid of the \Ha{} emission line,  shift the spectrum to its rest frame and divide by the peak \Ha{} flux. We remove the stellar continuum by subtracting a 4$^{\rm{th}}$ order polynomial fit and combine all spectra into a median stack. Our conclusions are unchanged if we sigma-clip the spectra before combining. The final stacked spectrum of each sample is shown in Figure \ref{fig:stacked_spectra}, where we also show a representative spectrum from an individual galaxy.  

We perform 10,000 bootstrap resamples to assess the uncertainties in each stack. If $N$ objects contribute to a stack, we randomly draw $N$ spectra from the sample (with replacement) and recombine them. The final error ``spectra" are estimated by taking the standard deviation of the bootstrap samples at each wavelength. We then measure the emission lines in the same manner as for individual galaxies (see Section \ref{sec:Ha_line_maps}), with measurement uncertainties estimated using 10,000 bootstrap resamples of each stacked spectrum.

\subsection{Gas-phase metallicities}
\label{sec:emline_results}

\begin{table}
\centering
\caption[Emission line ratios and derived quantities for the stacked spectra]{Emission line ratios and derived quantities for the stacked spectra. We fix the maximum value of the  {[\ion{S}{ii}]}$\lambda 6716$ / {[\ion{S}{ii}]}$\lambda 6731$ ratio to be 1.44 (see Section \ref{sec:EM_line_ratios}), and as such the upper uncertainty on this ratio in the field stacked spectrum is 0.00. The solar oxygen abundance is $12+\log(\rm{O/H}) = 8.69$ \protect\citep{Asplund:2009}.}
\begin{tabular}{lcc}
\toprule
  & Cluster  & Field  \\
\midrule
	{[\ion{N}{ii}]}$\lambda6584/$\Ha{} & $0.26\pm0.03$ & $0.25\pm0.02$ \\
       {[\ion{S}{ii}]}$\lambda 6716$ / [\ion{S}{ii}]$\lambda 6731$ & $1.18\pm0.17$  & $1.43^{+0.01}_{-0.02}$  \\
       {[\ion{S}{ii}]}$\lambda6716,6731/$\Ha{} & $0.22\pm0.04$ & $0.25\pm0.02$  \\
       {[\ion{N}{ii}]}$\lambda6584$/{[\ion{S}{ii}]}$\lambda6716,6731$ & $1.20\pm0.28$ & $1.01\pm0.15$ \\
       $12+\log(\rm{O/H})$ & $8.57\pm0.02$ & $8.56\pm0.02$ \\
       $n_e\,(\rm{cm}^{-3})$ & $126^{+182}_{-116}$ & $<10^{+28}_{-0}$ \\
\bottomrule
\end{tabular}
\label{tab:emission_line_ratios}
\end{table}

The stacked spectra of the two samples are shown in Figure \ref{fig:stacked_spectra}. It is clear that the mass-matched field and cluster galaxies show very similar average spectra. This adds to the findings of a number of other studies which show that the environment a galaxy resides in plays only a minor role in setting the conditions of its interstellar medium \citep[e.g.][]{Mouhcine:2007, Cooper:2008, Pilyugin:2017, Wu:2017}. 

A number of characteristics of star-forming regions can be investigated using emission-line fluxes and line ratios. Firstly, we measure the gas-phase metallicity, $12+\log(\rm{O/H})$, of the galaxies in our sample. A number of methods exist to convert emission-line measurements to metallicities, although it is well known that large discrepancies exist between metallicities estimated using different methods \citep[e.g.][]{Pilyugin:2001,Liang:2007, Kewley:2008}. Here, we derive the gas-phase metallicity using the ratio [\ion{N}{ii}]$\lambda6584/$\Ha{} and the polynomial conversion of \cite{Pettini:2004}: 

\begin{equation}
\label{eqn:mass_metallicity}
12+\log(\rm{O/H})=9.37 + 2.03N +1.26N^{2} + 0.32N^{3},
\end{equation}

\begin{figure*}
\begin{center}
\includegraphics[width=0.96\textwidth]{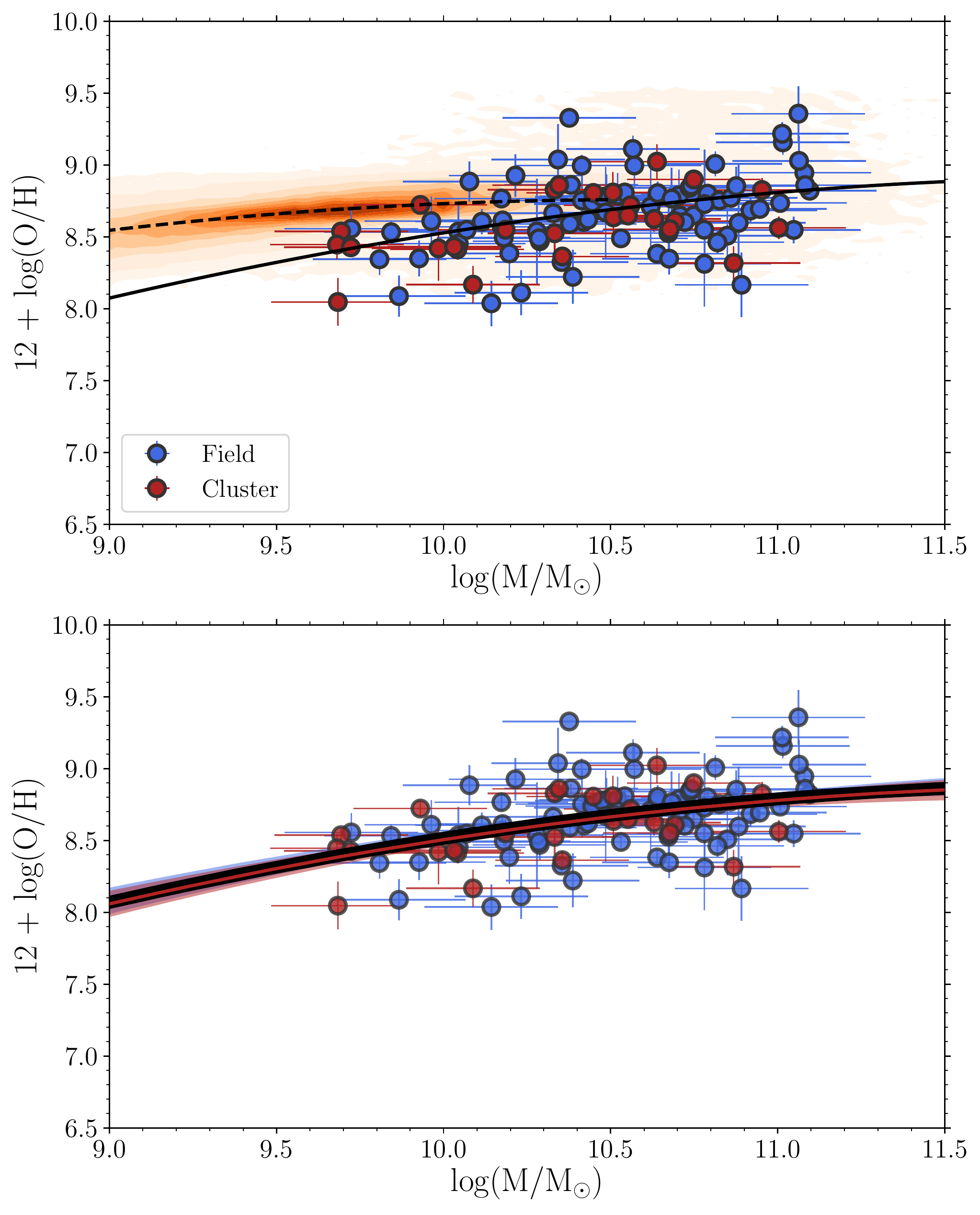}
\caption{Top: Mass-metallicity (MZ) relation for K-CLASH star-forming galaxies. We use the polynomial conversion of \protect\cite{Pettini:2004} to infer 12 + $\log$(O/H) from measurements of [\ion{N}{ii}]/\Ha{}. Cluster galaxies are shown in red with field galaxies in blue. The orange contours show local star-forming galaxies from the 12$^{\rm{th}}$ data release of the Sloan Digital Sky Survey, with emission line measurements from \protect\cite{Thomas:2013} and stellar masses from \protect\cite{Maraston:2009}. Solid and dashed lines show a polynomial fit to the K-CLASH and SDSS galaxies of the form of Equation \ref{eqtn:M_Z_relation}, following \protect\cite{Maiolino:2008}. Bottom: The K-CLASH sample split by environment. Red and blue lines show best-fit polynomials from Equation \ref{eqtn:M_Z_relation} to the cluster and field galaxies, respectively. These fits are derived from a hierarchical Bayesian model: see Equation \ref{eqtn:hierarchical_model} for details. Shaded regions represent the one $\sigma$ uncertainties around the fits. We note that we have already removed galaxies with a large [\ion{N}{ii}]/\Ha{} ratio (and, therefore, large 12 + $\log$(O/H) values) in their central spectrum, since they are likely to contain an AGN (see Section \ref{sec:AGN_removal}).}
\label{fig:M_Z}
\end{center}
\end{figure*}

\noindent where $N\equiv\log_{10}($[\ion{N}{ii}]$\lambda6584/$\Ha{}).We recall that we have already removed all galaxies with large [\ion{N}{ii}]$\lambda6584/$\Ha{} from our sample in an effort to remove galaxies containing an AGN. Whilst it is true that studies have shown that galaxies with the same [\ion{N}{ii}]$\lambda6584/$\Ha{} ratio can have different [\ion{O}{III}]/Ha{} ratios (and therefore different metalliicties: e.g. \citealt{Maier:2016}), with only the \Ha{} and \ion{N}{ii} emission lines available to us this conversion is the only one we are able to use. It does allow for easy comparison to gas-phase metallicity measurements at high-redshift, however, as many studies also use the [\ion{N}{ii}]$\lambda6584/$\Ha{} ratio \citep[e.g.][]{Swinbank:2012b, Stott:2014, Wuyts:2016, Magdis:2016}. 

We measure an [\ion{N}{ii}]$\lambda$6584/\Ha{} ratio of 0.26$\pm$0.03 for the cluster stacked spectrum and 0.25$\pm$0.02 for the mass-matched field stacked spectrum. These results are summarised in Table \ref{tab:emission_line_ratios}, and correspond to $12+\log(\rm{O/H})$= $8.57\pm0.02$ and $8.56\pm0.02$ respectively. For reference, the solar oxygen abundance is $12+\log(\rm{O/H}) = 8.69$ \citep{Asplund:2009}.

We also use Equation \ref{eqn:mass_metallicity} to construct the mass-metallicity (MZ) relation for individual galaxies \citep[see e.g.][]{Lequeux:1979, Tremonti:2004}. This is shown in the top panel of Figure \ref{fig:M_Z}.  Contours show the local MZ relation derived from 236,114 galaxies from the 12th data release of the Sloan Digital Sky Survey \citep[SDSS;][]{SDSS_DR12}.  Emission line measurements are from \cite{Thomas:2013}, with stellar masses estimated using the technique of \cite{Maraston:2009} assuming a Kroupa Initial Mass Function \citep[IMF;][]{Kroupa}. We inferred metallicities for the SDSS galaxies again using Equation \ref{eqn:mass_metallicity}. The average redshift of these objects is 0.06, with galaxies selected to be in the ``star-forming'' region of the BPT diagram.

Following \cite{Maiolino:2008}, we use a relation of the form 

\begin{equation}
\label{eqtn:M_Z_relation}
12 + \log(\textrm{O/H}) = -0.0864\left(\log \left(\frac{M}{\mathrm{M}_{\odot}}\right)-M_{0}\right)^{2}+K_{0},
\end{equation}

\noindent with free parameters $M_0$ and $K_0$; $K_0$ corresponds to the metallicity of a galaxy with mass $M_0$. We again perform the regression using \texttt{Stan}, incorporating uncertainties in the $x$ and $y$ directions and intrinsic scatter around the relation. We place Gaussian priors of $\mathcal{N}(10, 2)$ on $M_0$ and $K_0$ and a "half-normal" prior\footnote{defined as a normal distribution for positive values of the dependent variable and zero otherwise.} of $\mathcal{N}(0, 1)$ on the intrinsic scatter parameter $\sigma$. For the entire K-CLASH sample, we find $M_0 = 12.12\pm0.34$ and $K_0=8.92\pm0.10$, with an intrinsic scatter of $\sigma=0.18\pm0.02$ dex.

In the bottom panel of Figure \ref{fig:M_Z}, we split our sample by environment and fit a Bayesian hierarchical model to the cluster and field samples. Rather than fitting to the two populations individually, we describe the unknown model parameters of the cluster and field populations as being drawn from shared prior distributions, with these prior distributions themselves described by shared hyper-parameters that are also estimated during the fitting. This allows for inference on the unknown parameters in the cluster and field samples separately, whilst also resulting in tighter constraints on their measurement; both populations can borrow strength from one another by influencing the shared hyper-parameter posterior distributions. In this way, hierarchical modelling is the best compromise between fitting to the cluster and field samples completely independently (resulting in larger uncertainties on $M_0$ and $K_0$ for both populations) and combining all galaxies together to derive single values of $M_0$ and $K_0$ (which prevents us inferring any differences between the two samples). An introduction to hierarchical models can be found in \cite{BDA3}, with some recent discussion and examples of their use in astronomy in e.g. \cite{Lieu:2017},  \cite{Sharma:2017},  \cite{Thrane:2019},  \cite{Grumitt:2019}. We fully describe our model below. In the following context, the symbol $\sim$ means "is distributed according to", e.g. $\alpha \sim \mathcal{N}(0, 2)$ means that the parameter $\alpha$ is distributed according to a normal distribution with mean 0 and standard-deviation 2.

\begin{align}
\label{eqtn:hierarchical_model}
\begin{split}
i &= 1...N_{\mathrm{galaxies}}\\
j &= \mathrm{field~or~cluster}\\
\midrule
\alpha &\sim \mathcal{N}(0, 2)\\
\beta &\sim \mathrm{Inv-gamma}(2, 0.1)\\
\gamma &\sim \mathcal{N}(0, 2)\\
\delta &\sim \mathrm{Inv-gamma}(2, 0.1)\\
\tau &\sim \mathrm{Inv-gamma}(2, 0.1)\\
\midrule
M_{0, j} &\sim \mathcal{N}(\alpha, \beta)\\
K_{0, j} &\sim \mathcal{N}(\gamma, \delta)\\
\sigma_{j} &\sim \mathrm{Half-}\mathcal{N}(0, \tau)\\
\midrule
M_{\rm{true}, i} &\sim \mathcal{N}(M_{\rm{obs}, i}, \sigma_{M, i})\\
y_{\rm{true}, i} &\sim \mathcal{N}(y_{\rm{obs}, i}, \sigma_{y, i})\\
\theta_i &= -0.0864\left(\log \frac{M_{\rm{true, i}}}{\mathrm{M}_{\odot}} -\frac{M_{\rm{0, i}}}{\mathrm{M}_{\odot}} \right)^{2}+K_{0, j}\\ 
y_{\rm{true}, i} &\sim \mathcal{N}(\theta_{i}, \sigma_{j})
\end{split}
\end{align}

This model should be interpreted as follows. The index $i$ labels individual galaxies, and runs from 1 to the total number of objects in our sample.   For each galaxy, the quantities we want to relate are its true value of $12 + \log(\mathrm{O/H})$ and its true value of stellar mass. We denote the true gas-phase metallicities of our samples to be $y_{\rm{true}, i}$. This vector $y_{\rm{true}, i}$ is a "latent" variable, in that we do not (and cannot) observe it directly. Instead, we only have uncertain measurements of our galaxies' gas-phase metallicities, which we denote $y_{\rm{obs}, i}$. We relate $y_{\rm{true}, i}$ to $y_{\rm{obs}, i}$ using a series of Gaussian distributions. These distributions are centred on $y_{\rm{obs}, i}$ and have standard deviations given by the measurement uncertainties on $y_{\rm{obs}, i}$, $\sigma_{y, i}$. The same is true for each galaxy's stellar mass: we relate our noisy observations ($M_{\rm{obs}, i}$) to each galaxy's true stellar mass ($M_{\rm{true}, i}$) using a series of Gaussian distributions with standard deviations $ \sigma_{M, i}$.

The quantities we wish to infer, $M_0$, $K_0$ and the intrinsic scatter $\sigma$, may take different values for the cluster and field samples. We use the index $j$ to show this; $j$ can take the values ``field'' or ``cluster'', depending on whether galaxy $i$ is in the field or cluster sample. For each galaxy, we use Equation \ref{eqtn:M_Z_relation} to infer a value of gas-phase metallicity from $M_0$, $K_0$ and $M_{\rm{true}, i}$. We then describe $y_{\rm{true}, i}$ as being distributed as a series of Gaussians centred on these value of gas-phase metallicity, with standard deviation $\sigma_j$. 

We place Gaussian priors (denoted $\mathcal{N}$) on  $M_{0, j}$ and $K_{0, j}$, and a half-Gaussian prior on $\sigma_j$. The parameters $\alpha, \beta, \gamma, \delta$ and $\tau$ are hyper-parameters. We place Gaussian priors on $\alpha$ and $\gamma$ (which describe the "location" of the priors on $M_{0, j}$ and $K_{0, j}$) and inverse gamma priors (denoted $\mathrm{Inv-gamma}$ ) on $\beta, \gamma$ and $\tau$ (which denote the "width" or "scale" of the priors on $M_{0, j}$, $K_{0, j}$ and $\sigma_j$). We choose an inverse gamma prior for these quantities to ensure they remain positive. During the fitting process, we took the standard modelling step of centring our observations around zero by subtracting their mean value. The model was fit using \texttt{Stan}, with the ``maximum tree-depth'' parameter set to 20. 

We find that the two samples lie on indistinguishable mass-metallicity relations: for the cluster sample $M_0 = 12.07\pm0.34$, $K_0=8.89\pm0.11$ and $\sigma=0.15\pm0.03$ dex;  for the field sample $M_0 = 12.07\pm0.34$, $K_0=8.91\pm0.09$ and $\sigma=0.19\pm0.02$ dex. We also note that our conclusions remain unchanged if we perform a standard fit to the field and cluster populations separately, instead of using the hierarchical model outlined above. 

The fact that the field and cluster MZ relations are the same is in agreement with \cite{Maier:2016}, who found the difference between the MZ relations of field and cluster galaxies (in another CLASH cluster at $z\approx0.4$) to be less than 0.1 dex. Similarly, for local galaxies, \cite{Mouhcine:2007} found only small differences between the gas-phase metallicity of galaxies with masses greater than $10^{9.5}$ \Msun{} as a function of environmental density. On the other hand, after removing the trend between environment, colour and luminosity, \cite{Cooper:2008} found a weak but significant trend between metallicity and environment, with more metal-rich galaxies residing in higher density environments. \cite{Ellison:2009} also found an elevation of 0.04 dex in metallicity between a sample of 1318 cluster galaxies and a matching sample of field galaxies, but also a stronger trend between \textit{local} density, rather than simply cluster membership, and metallicity. Finally, \cite{Gupta:2016} studied the MZ relation in two CLASH clusters at $z\sim0.35$, finding that the relation of galaxies residing in RX J1532+30 is consistent with their local comparison sample whilst the relation of galaxies in MACS J1115+01 is enhanced by 0.2 dex.

\subsection{Residuals around the Mass-Metallicity relation}

\label{sec:MZ_residuals}

\begin{figure*}
\begin{center}
\includegraphics[width=\textwidth]{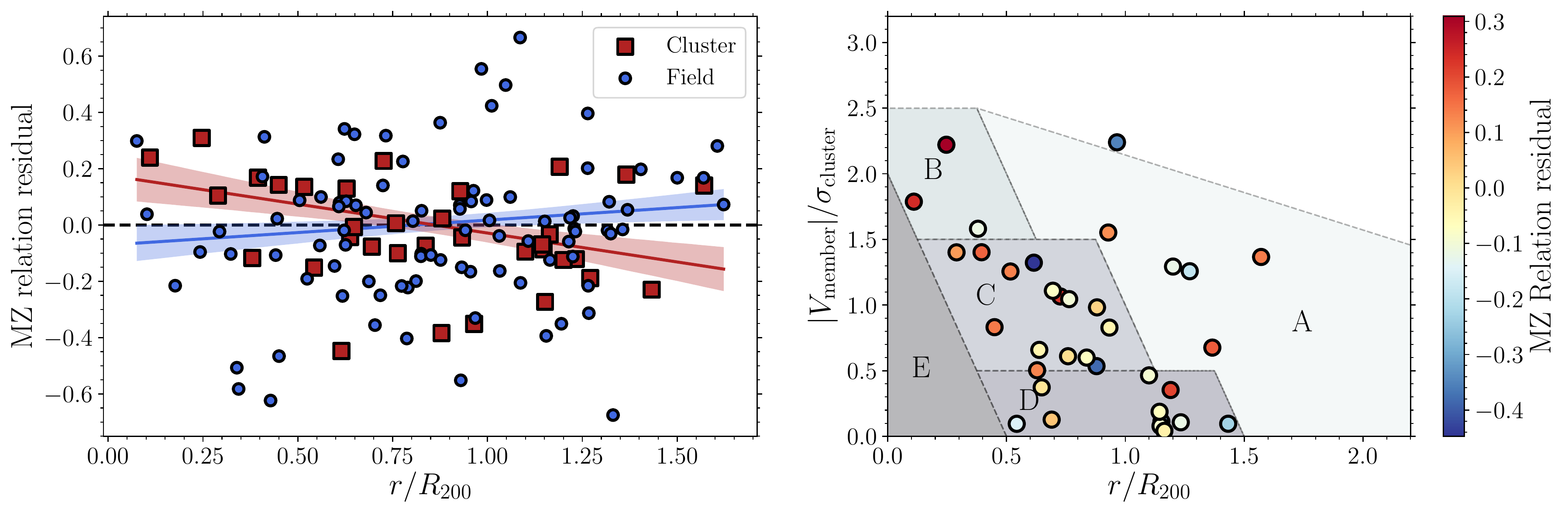}
\caption[]{Correlation between a galaxy's residual around the mass-metallicity relation and its location within its galaxy cluster. Left panel: we plot the MZ relation residuals against projected radii (scaled by the appropriate cluster $R_{200}$ value) for the field (blue) and cluster (red) samples. There is no significant correlation between the MZ relation residuals and stellar mass or SFR for either sample. Right panel: for galaxies in the  cluster sample, we show their locations in cluster phase-space. Each point is coloured by its residual around the MZ relation. We also use the simulations of \protect\cite{Rhee:2017} to show the regions in phase-space which roughly correspond to the locations where first infalling galaxies (A) and recent/intermediate infallers (B and C) reside. Region D is generally made up of a combination of intermediate and ancient infallers, as well as some "backsplash" galaxies, whilst E is dominated by ancient infallers. See Section \ref{sec:MZ_residuals} for details. }
\label{fig:MZ_residual_with_cluster_position}
\end{center}
\end{figure*}

Whilst we do not detect a difference between the mass-metallicity relation of galaxies residing in high-density environments (our cluster sample) and lower-density environments (our field sample), a number of studies have documented a correlation between a galaxy's location in cluster phase space and its metal content. \cite{Maier:2016} showed that the fraction of accreted galaxies that are metal-rich is higher than their sample of infalling galaxies, and these accreted galaxies have higher metallicities than predicted from models assuming a constant supply of inflowing pristine gas. Similarly, \cite{Gupta:2016} measured a correlation between the residuals around the MZ relation and cluster-centric distances in one of the two massive CLASH clusters they studied (although they found no correlation in the other). In the local Universe,  \cite{Pilyugin:2017} found that galaxies in the densest environments have a median increase in Oxygen abundance of 0.05 dex with respect to the MZ relation, whilst \cite{Wu:2017} also showed that the median MZ relation residual is a weak function of environment, with a primary dependence on stellar mass.

To investigate these effects in our own samples, we fit a linear model to the residuals around the MZ relation for the cluster and field samples separately.  We define $\Delta(\mathrm{O/H})$ for each galaxy to be its gas-phase metallicity measurement minus the metallicity value from the MZ relation at the galaxy's stellar mass.  Our model is  of the form

\begin{equation}
\Delta(\mathrm{O/H}) = \alpha + \beta_1 (r/R_{200}) + \beta_2 \log(M_{*}/M_{\odot}) + \beta_3 \mathrm{SFR}/(M_{\odot}\mathrm{yr}^{-1})
\end{equation}

\noindent which includes the projected distance from the cluster centre, $r$, (scaled by the $R_{200}$ value of the appropriate cluster), stellar mass and star formation rate as explanatory variables. We again use \texttt{Stan} to infer the posterior probability distribution of each coefficient, finding that the $\Delta(\mathrm{O/H})$ has no dependence on $M_*$ or SFR for the cluster or field samples. The only correlation coefficient more than one standard deviation away from zero occurs with projected distance for the cluster sample:  $\beta_{1, \mathrm{cluster}} = -0.21 \pm 0.08$, significant at the 2.6$\sigma$ level. As expected, the field sample coefficient is consistent with zero: $\beta_{1, \mathrm{field}} = 0.04 \pm 0.04$. We show the correlation between projected distance and the MZ relation residual in the left-hand panel of Figure \ref{fig:MZ_residual_with_cluster_position}. 

The right-hand panel of Figure \ref{fig:MZ_residual_with_cluster_position} plots each galaxy in the cluster sample in phase-space, coloured by the galaxy's residual above or below the MZ relation. We also show regions in phase space from the simulations of \cite{Rhee:2017}, labelled A through E. Region A contains the largest fraction of "first infallers" into the cluster, whilst "recent" infallers (galaxies which have fallen into the cluster 0 - 3 Gyr ago) and intermediate infallers (3 - 6.5 Gyrs ago) tend to be found in regions B and C. Region E, containing "ancient" infallers (accreted > 6.5 Gyrs ago), is underpopulated in our sample, showing that these galaxies are no longer visible in \Ha. Region D tends to contain a combination of intermediate and ancient infallers, as well as a population of "backsplash" galaxies. We make further comment on these timescales in Section \ref{sec:Discussion}. 

\subsection{ISM conditions}
\label{sec:EM_line_ratios}

\begin{figure}
\begin{center}
\includegraphics[width=0.5\textwidth]{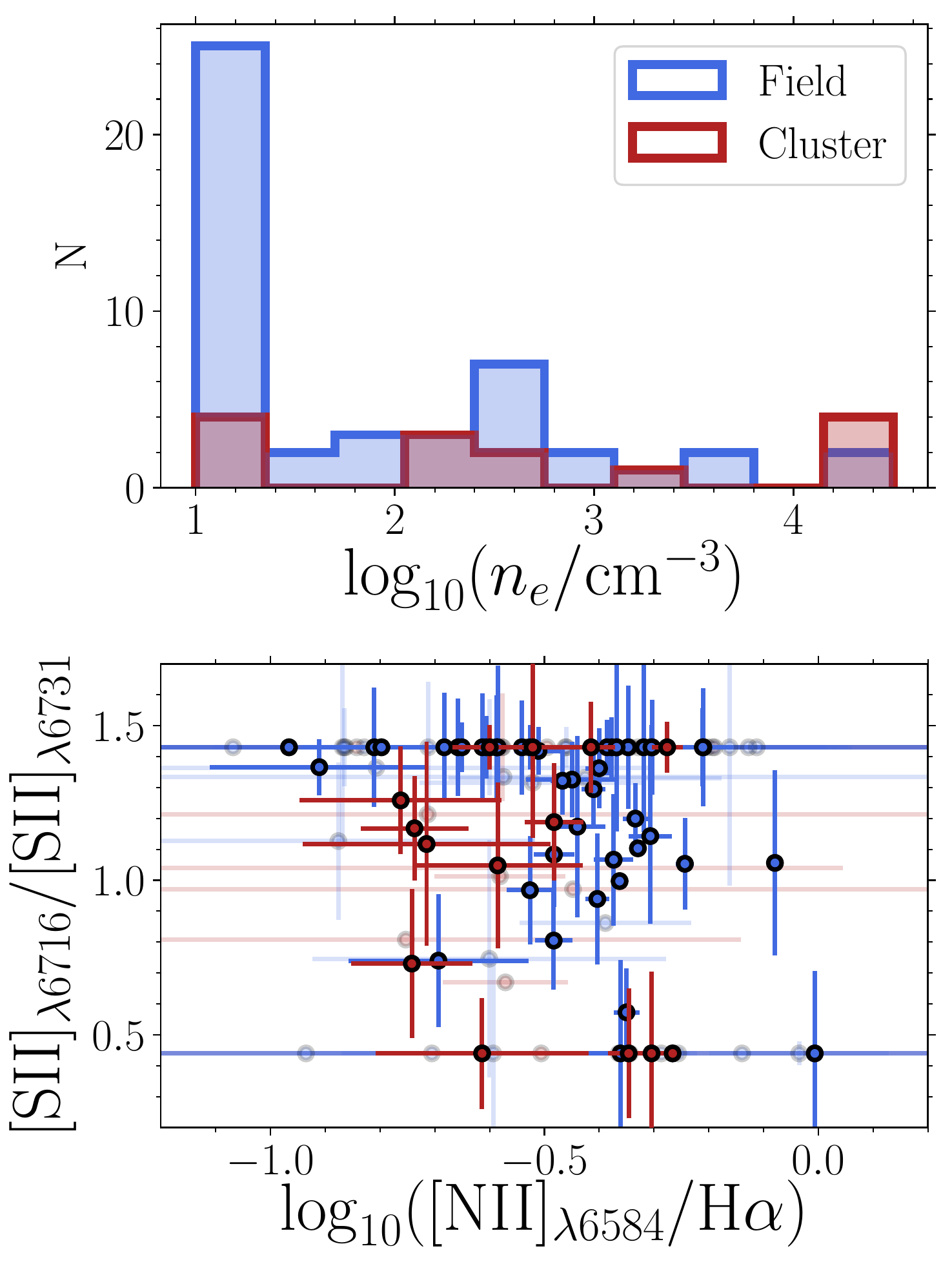}
\caption[ {[\ion{N}{ii}]}/\Ha{} against {[\ion{S}{ii}]} ratio]{Top: A histogram of the electron number density in the mass-matched field (blue) and cluster (red) galaxy samples, estimated from the  [\ion{S}{ii}]$\lambda 6716$/ [\ion{S}{ii}]$\lambda 6731$ line ratio and the calibration of \protect\cite{Proxauf:2014}. Bottom: The [\ion{S}{ii}]$\lambda 6716$/[\ion{S}{ii}]$\lambda 6731$ ratio against $\log_{10}($[\ion{N}{ii}]$_{ \lambda6584}/$\Ha{}). Each solid point has [\ion{S}{ii}] and [\ion{N}{ii}] S/N greater than 3, whilst faded points have S/N less than 3 in at least one line. We limit values of [\ion{S}{ii}]$\lambda 6716$/ [\ion{S}{ii}]$\lambda 6731$ to be between 0.44 and 1.44, the maximal values allowed by atomic physics.}
\label{fig:electron_density}
\end{center}
\end{figure}

The ratio [\ion{S}{ii}]$\lambda 6716$ / [\ion{S}{ii}]$\lambda 6731$ is a well-known electron number density diagnostic tool \citep[e.g.][]{Osterbrock:2006, Proxauf:2014}. The line ratio in the very low ($n_{\mathrm{e}}<10$ cm$^{-3}$) and high ($n_{\mathrm{e}}>10^{4}$ cm$^{-3}$) electron number density limits are 1.44 and 0.44, respectively.  Assuming an electron temperature of 10,000 K and using the empirical calibration of \cite{Proxauf:2014}, we find that the average electron number density in the stacked spectra from the cluster sample and mass matched field sample are $n_{e}=126^{+183}_{-116}$ cm$^{-3}$ and $<10^{+28}_{-0}$ cm$^{-3}$ respectively. 

Figure \ref{fig:electron_density} shows the derived electron number densities and a comparison between the [\ion{S}{ii}]$\lambda 6716$/[\ion{S}{ii}]$\lambda 6731$ and [\ion{N}{ii}]$_{ \lambda6584}$/\Ha{} ratios. Among the individual objects, the distribution of electron number densities for the field and cluster galaxies values are similar, although interestingly we do observe proportionally fewer cluster galaxies with a small [\ion{S}{ii}]$\lambda 6716$/[\ion{S}{ii}]$\lambda 6731$ ratio compared to the field sample.

These comparable electron number densities across both environments are in agreement with \cite{Kewley:2016}, who found no difference between the electron number densities of a sample of 13 galaxies in a $z=2.1$ proto-cluster and a number of $z\approx2$ field galaxies. It should be noted, however, that the general properties of  $z\approx2$ galaxies and the galaxy population at $0.3 < z < 0.6$ are very different- as are the environmental conditions in high-redshift proto-clusters and the massive  intermediate-redshift clusters studied in this work.

On the other hand, some studies have found an environmental dependence of the electron number density. \cite{Darvish:2015} observed galaxies residing in large-scale filamentary structures at a similar redshift to our sample ($z\approx0.5$), finding significantly smaller electron number densities than those in a similar sample residing in the field. Similarly, \cite{Sobral:2015} studied a merging cluster at $z\approx0.2$ and found that the cluster galaxies have electron number densities lower than that of field objects with similar properties. Conversely, at higher redshift \cite{Harshan:2020} found that galaxies in a $z=1.62$ proto-cluster have \textit{higher} electron number densities than a those of a similar field sample, significant at the 2.6$\sigma$ level.

\section{Discussion}
\label{sec:Discussion}

This work has two main conclusions. Firstly, the average \Ha{} to continuum size ratio (\sizeratio{}) of star-forming cluster galaxies is smaller than that of star-forming field galaxies which are matched in mass (Section \ref{sec:Ha_Results}). Secondly, the emission line ratios of the integrated spectra of cluster and field galaxies lead to identical mass-metallicity (MZ) relations.  For the cluster sample, however, the residuals around the MZ relation are correlated with a galaxy's cluster-centric distance, with galaxies closer to the centre of their cluster preferentially scattering to higher gas-phase metallicities by up to $\sim0.2-0.3$ dex above the relation (Section \ref{sec:MZ_residuals}). We also find a number of secondary conclusions: that galaxies in the cluster sample have an average \Ha{} surface brightness within a 0\farcs6 aperture which is marginally fainter than those in the field sample by 0.06 dex; and that the ISM conditions of the galaxies in the two samples are similar, with comparable electron densities (but tentative evidence that $n_e$ in the cluster galaxies is larger).

\subsection{Environmental effects}

A number of environmental processes have been suggested to quench galaxy cluster members, with the most commonly proposed being "strangulation" and "ram-pressure stripping".

Strangulation occurs when a galaxy's supply of cold gas residing in its halo is removed. In the absence of a supply of cold gas, galaxies continue to form stars until they run out of the fuel residing in their discs \citep[e.g.][]{Larson:1980, Peng:2015}. This leads to an overall reduction of the star-formation rate and \Ha{} flux, but also an increase of the gas-phase metallicity (as the ISM is no longer diluted by an inflow of low-metallicity material). \cite{Maier:2016}, for example, found strangulation to be consistent with their measurements of the chemical enrichment of the galaxies in a CLASH cluster at $z\approx0.4$ (and see also \citealt{Maier:2019a, Maier:2019b, Ciocan:2020} for studies at lower and higher redshift). 

We use the ``bathtub'' chemical evolution model of \cite{Lilly:2013}, \cite{Peng:2014} and \cite{Peng:2015} to study the evolution of the metallicity and central surface brightness of a galaxy whose halo of cold gas has been removed. Following this model, at a time $t$ after the onset of disc strangulation ($t_q$) the galaxy's increase in metallicity ($\Delta \log[Z(t)]$) is:

\begin{equation}
\Delta \log[Z(t)] = \log\left(1+ \frac{y \varepsilon t}{Z(t_q)}\right)
\end{equation} 

\noindent where $y$ is the average metal yield per stellar generation and $\varepsilon$ is the star-formation efficiency. 

Furthermore, we can model the surface brightness within a 0\farcs6 diameter aperture.  Firstly, we assume that the gas follows an exponential surface brightness distribution with a scale length 1\arcsec (approximately that found in the cluster and field galaxies). Secondly, we use the stellar masses derived in \citetalias{Tiley:2020}, assume a gas fraction and use the assumed exponential scale length to find a central gas surface density. Under the assumption that the star formation rate is related to the total gas mass ($\mathrm{SFR}=\varepsilon M_{g}$), we use the conversion of \cite{Hao:2011} and \cite{Murphy:2011} to derive an observed \Ha{} flux from a star-formation rate (including a dust extinction of $A_V=0.5$ magnitudes, the average value found in the K-CLASH sample; see \citetalias{Tiley:2020}).  We then integrate the exponential disc profile within the aperture to obtain a surface brightness value. 

Over time, as the galaxy consumes its gas, the gas mass (and hence the SFR and \Ha{} flux) will decrease exponentially according to Equation 15 of \citealt{Peng:2014}: $\mathrm{SFR} \propto \exp(-\varepsilon(1-R)t)$. Here, $R$ is the fraction of mass of newly-formed stars which is (instantaneously) returned to the ISM via supernovae and stellar winds. 

Following \cite{Lilly:2013}, we use values of $R=0.4$ and $y = 0.016$ (i.e. $y \approx 9$ in units of $12 + \log(\mathrm{O/H})$). We also use a gas fraction of 1, a star-formation efficiency of 0.1 Gyr$^{-1}$ and a solar metallicity at the start of strangulation ($Z(t_q) = 0.0134$; \citealt{Asplund:2009}). Finally, based on the simulations of \cite{Rhee:2017} and the locations of our targets in Figure \ref{fig:MZ_residual_with_cluster_position}, we assume an infall time of both 1 and 3 Gyrs (i.e. 1 or 3 Gyrs since strangulation). We then derive the following: simulated surface brightnesses which match those shown in Figure \ref{fig:surface_brightness}; a decrease in central surface brightness after 1 (3) Gyrs of disc strangulation to be $\approx0.05~(0.15)$ dex; and a gas-phase metallicity increase of $\approx 0.1~(0.2)$ dex. These values are in agreement with our findings in this work, although we caution that this analysis is approximate in nature; we want to show that a simple strangulation model can explain our measurements, rather than perform a full quantitative analysis of our results. 

A simple toy model of strangulation can therefore account for both the $0.06$ dex decrease in central surface brightness as well as the $\approx0.2$ dex scatter to higher metallicities in the cluster sample galaxies, in a timescale which matches the cluster infall times implied by the simulations of \cite{Rhee:2017}. However, this simple model can \textit{not} account for the decrease in half-light radii measured in Section \ref{sec:Ha_Results}; if gas is being consumed throughout the galaxy disc, to a first approximation the entire disc would become less star-forming but its scale length would remain unchanged \citep[e.g.][]{Bekki:2002, Boselli:2006b}.  We note that our modelling in Section \ref{sec:Ha_Results} measured the \textit{intrinsic} half-light radii of our targets, rather than the observed half-light radii. As such, even if a star-forming disc appears to become smaller (as its outer regions fell below our detection threshold), we will still recover the same half-light radius. Since we do in fact measure slightly smaller average \sizeratio{} in the cluster sample galaxies, another process must be at play.

We therefore conclude that our galaxy sample is also being affected by ram-pressure stripping. Ram-pressure stripping occurs when the pressure of the intra-cluster medium (ICM) "wind" experienced by a galaxy (due to its motion through the ICM) exceeds the galaxy's gravitational restoring force and begins to remove material from its outskirts \citep{Gunn:1972}. It has also been shown to reduce the size of \Ha{} discs. Of the many hydrodynamical simulations of ram-pressure stripping available in the literature, the study of \cite{Bekki:2014} is the most appropriate to compare to our work. They used hydrodynamical simulations of ram pressure to investigate the ratios of \Ha{} to optical disc scale lengths of galaxies passing through dense environments. They found that whilst  the precise evolution of  $r_{e, {\rm{H}\alpha}}/r_{e, \mathrm{optical} }$ for individual star-forming galaxies in clusters depends sensitively on the cluster halo mass and galaxy disc inclination with respect to the cluster core,  in general ram-pressure stripping reduces the $r_{e, {\rm{H}\alpha}}/r_{e, \mathrm{optical} }$ ratio in disc galaxies in massive clusters, which matches the findings in this work. They also found that the central star formation of these galaxies can be moderately enhanced (during pericentre passage, primarily for edge-on systems), suppressed or completely quenched (both after pericentre passage). 

We note that, by definition, the process of ram-pressure stripping also encompasses the effects of disc strangulation, and therefore our previous calculations regarding the increase in metallicity and reduction in surface brightness are still valid. We also note that it is only through the use of integral-field observations, which allow us to make measurements of the extent of \Ha{} discs and gas-phase metallicities at the same time, that we have been able to come to this conclusion. We therefore strongly advocate the use of spatially resolved spectroscopy in future studies of environmental quenching processes. 

Local studies have recently unveiled the complexity of galaxies undergoing gas-stripping processes. The GASP project (GAs Stripping Phenomena in galaxies with MUSE; \citealt{Poggianti:2017} and references therein) studies 114 nearby galaxies in group and cluster environments which show evidence of recent stripping by ram pressure or turbulent processes. Truncated gas discs are common in the galaxies published so far, with most observations also showing evidence of spectacular tails of ionised gas \citep{Poggianti:2017,Gullieuszik:2017, Moretti:2018}, although these are not ubiquitous \citep{Fritz:2017}. The 3$\sigma$ limiting surface brightness of the GASP observations is $2.5\times10^{-18}$ erg s$^{-1}$ cm$^{-2}$ arcsec$^{-1}$ \citep{Poggianti:2017}, with the tidal tails having surface brightnesses of $\lesssim1\times10^{-16}$ erg s$^{-1}$ cm$^{-2}$ arcsec$^{-1}$ \citep[e.g.][]{Gullieuszik:2017}, below the average K-CLASH limiting surface brightness ($\approx1\times10^{-15}$ erg s$^{-1}$ cm$^{-2}$ arcsec$^{-1}$; see Section \ref{sec:Ha_line_maps}). We estimate that we would require nine hours on source to obtain a 3$\sigma$ detection of an \Ha{} emission line at 1 $\mu$m with a surface brightness of $1\times10^{-16}$ erg s$^{-1}$ cm$^{-2}$ arcsec$^{-1}$, 3.6 times longer than the typical K-CLASH galaxy observation.

\subsection{Electron number densities}

Despite the differences discussed above, the stacked spectra of cluster and field galaxies do not show a significant difference in electron number density measurements (although the uncertainty on these measurements are large). This finding would be in contrast to the work of \cite{Darvish:2015} and \cite{Sobral:2015}, who find smaller electron densities in higher density environments.

Our results could imply that the ram-pressure stripping has not yet directly impacted the ISM in the centres of our targets, where most of the emission line flux originates.  As a galaxy moves through the dense ICM, its gaseous halo and disc are compressed towards the cluster centre and stripped on the trailing edge. Physically, one might expect to see a variation of the gas density between the leading and trailing edges of the object, which could be evident in the ratio of  the [\ion{S}{ii}] doublet lines. Spatially resolved maps of the [\ion{S}{ii}] line ratio have been studied in local AGN, ultra luminous infrared galaxies (ULIRGs) and starburst galaxies  \citep[e.g.][]{Bennert:2006, Sharp:2010, Westmoquette:2011, Kakkad:2018}, but not for objects undergoing ram-pressure stripping, for which the GASP project provides an excellent data set.

\section{Conclusions}
\label{sec:Conclusions}
Using IFU observations from the K-CLASH survey \citep{Tiley:2020}, we have studied the effect of environment on star-forming galaxies in 4 CLASH clusters at $0.3<z<0.6$. We make comparisons to a mass-matched sample of galaxies residing in the field along nearby lines of sight at similar redshifts. We note that we cannot guarantee the purity of our cluster and field samples, as our simple cuts in projected radius and velocity do not account for the undoubtedly complex distribution of mass in each cluster. Our results, therefore, should be viewed as lower limits to the true differences between the cluster and field populations at these redshifts; any contamination (in either direction) will tend to homogenise our samples and reduce the diversity we find. 

Firstly, we infer the radial extent of ongoing star formation and older stellar populations by fitting exponential disc models to the \Ha{} surface brightness distributiomns and S\'ersic profiles to $R_\mathrm{c}$-band images. We have ensured that fitting the \Ha{} maps with more general S\'ersic profiles does not change our results. We then investigate the physical conditions of the ISM of galaxies in our sample by interpreting the emission line ratios measured in the integrated spectrum of each object, as well as stacking these spectra together to improve the signal-to-noise ratio and determine average properties. We summarise our conclusions below.

\begin{enumerate}
\item The average ratio of the half-light radius of the \Ha{} emission and the $R_\mathrm{c}$-band continuum emission (\sizeratio{}) across all galaxies is $1.14\pm0.06$, showing that star formation is generally taking place throughout stellar discs at these redshifts.
\item When separating by environment, we find an average \sizeratio{}=0.96$\pm$0.09 for galaxies in the cluster sample and $1.22\pm0.08$ for galaxies in the field sample. $\langle$\sizeratio$\rangle$ of the cluster galaxies is smaller than $\langle$\sizeratio$\rangle$ of the field galaxies at the 98.5\% confidence level.
\item The central surface brightnesses within a 0\farcs6 diameter aperture are $\approx0.05$ dex fainter for galaxies in the cluster sample than those in the field sample.
\item Using the conversion of \cite{Pettini:2004}, we measure a gas-phase metallicity for each object from the [\ion{N}{ii}]$\lambda6584$/\Ha{} ratio. Both the cluster and field galaxies follow indistinguishable mass-metallicity (MZ) relations. 
\item We do, however, see a correlation between a galaxy's residual around the MZ relation and its projected radius (for galaxies in the cluster sample). Galaxies which are residing closer to the centre of their parent cluster tend to be more metal enriched (by up to $\approx0.2-0.3$ dex more than expected, given their mass).
\item Using the ratio of the [\ion{S}{ii}]$\lambda6716$/[\ion{S}{ii}]$\lambda6731$ lines, we infer the electron number density, $n_e$, in each galaxy. The distribution of these values are broadly similar between the field and cluster samples, although we do find a smaller proportion of cluster galaxies with very low $n_e$ compared to the field sample. In contrast to previous studies, the stacked cluster spectrum and the stacked field spectrum do not show a significant difference in electron number density (although the large uncertainties prevent us from drawing strong conclusions from this result).  
\item We use the ``bathtub'' chemical evolution models of \cite{Lilly:2013} and \cite{Peng:2014} to show that removal of a galaxy's halo of cold-gas (i.e. "disc strangulation") can account for the fainter surface brightnesses and scatter to higher metallicities of galaxies in the cluster sample. However, since strangulation alone cannot explain the measured reduction in the intrinsic size of the \Ha{} discs, we conclude that ram-pressure stripping must also be affecting the outskirts of our targets.
\end{enumerate}

\section*{Acknowledgements}
We thank the anonymous referee for their comments which greatly improved this work.

This paper made use of the \texttt{astropy} \textsc{python} package \citep{astropy}, as well as the \texttt{matplotlib} plotting software \citep{matplotlib} and the scientific libraries \texttt{numpy} \citep{numpy}, \texttt{pandas} \citep{pandas} and \texttt{scipy} \citep{scipy}. 

This work was based on observations collected at the European Organisation for Astronomical Research in the Southern Hemisphere under ESO programme(s) 097.A-0397, 098.A-0224, 099.A-0207 and 0100.A-0296. Parts of this research were supported by the Australian Research Council Centre of Excellence for All Sky Astrophysics in 3 Dimensions (ASTRO 3D), through project number CE170100013. SPV acknowledges support from a doctoral studentship from the UK Science and Technology Facilities Council (STFC) grant ST/N504233/1.  ALT acknowledges support from a Forrest Research Foundation Fellowship, STFC (ST/L00075X/1 and ST/P000541/1), the ERC advanced Grant DUSTYGAL (321334), and an STFC Studentship. RLD acknowledges travel and computer grants from Christ Church, Oxford, and support from the Oxford Hintze Centre for Astrophysical Surveys, which is funded through generous support from the Hintze Family Charitable Foundation. MB was supported by the consolidated grants ‘Astrophysics at
Oxford’ ST/H002456/1 and ST/K00106X/1 from the UK Research Council.

\section*{Data Availability}
The data underlying this article are available in the article and in its online supplementary material.

\bibliographystyle{mnras}
\bibliography{bibliography}

\appendix

\section{Variations of the KMOS PSF}
\label{sec:KMOS_PSF_variation}

The KMOS instrument comprises three separate spectrographs, which disperse the light from arms 1--8, 9--16 and 17--24 respectively. It has been reported that the PSF can vary slightly between the spectrographs \citep[e.g.][]{Magdis:2016}, and so it is important to investigate whether any systematic differences exist which could impact our results. 

\begin{figure}
\begin{center}
\includegraphics[width=0.5\textwidth]{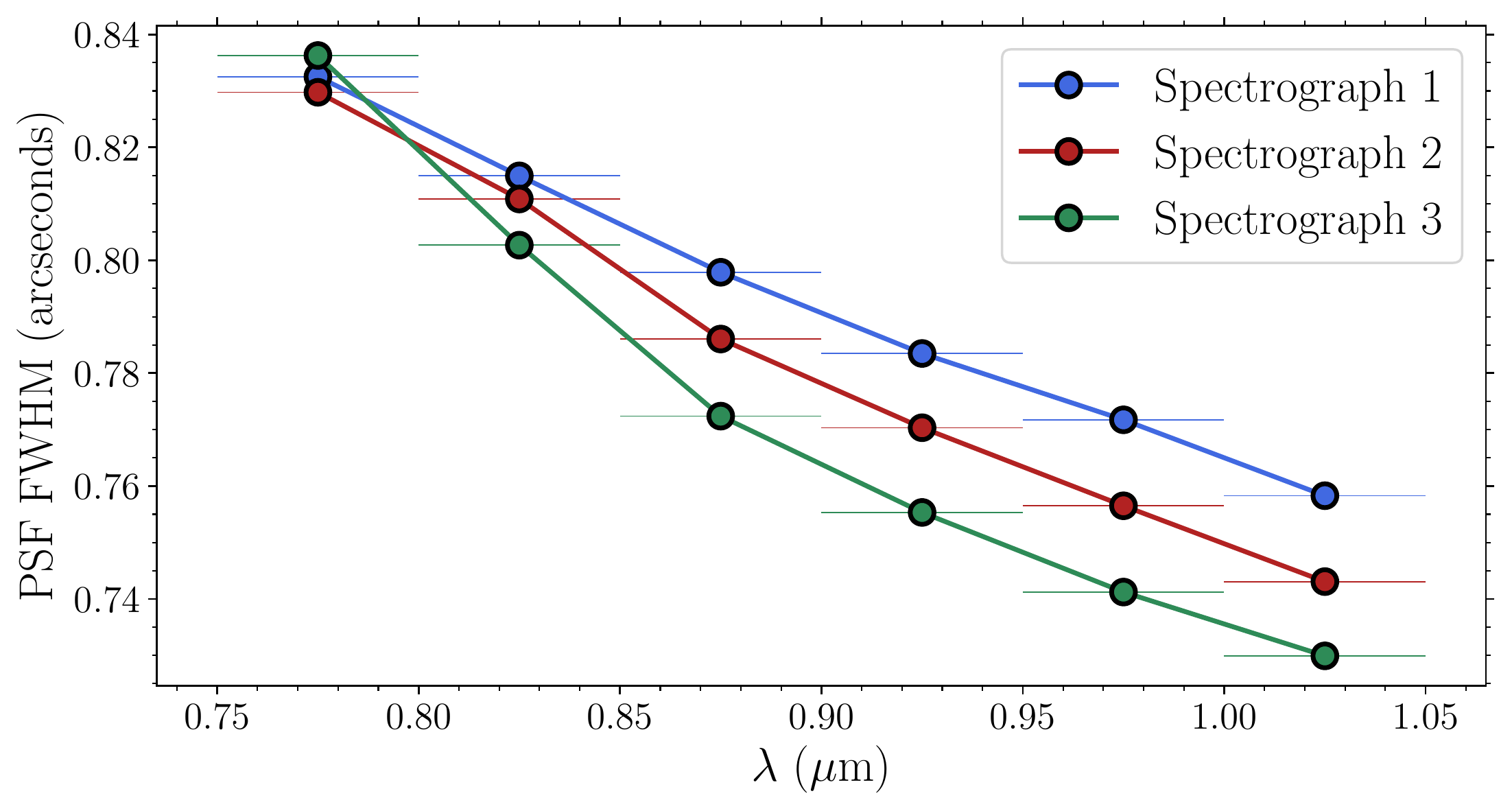}
\caption[]{The average FWHM of the PSF as a function of wavelength and colour-coded by KMOS spectrograph.  A two-dimensional image was created from each star observed during the K-CLASH survey by collapsing the full data-cube in a window of 0.05 $\mu$m in the wavelength direction. We then fit a two-dimensional Gaussian function to derive the FWHM of each observation, and average together all observations in the same spectrograph. This procedure is then repeated a further five times for increasing central wavelength. The FWHM of the PSF varies by around 0\farcs1 across the \textit{IZ} spectral range, and by 0.03\arcsec (at most) between spectrographs.}
\label{fig:spectrographs_PSF_variation_with_lambda}
\end{center}
\end{figure}

Firstly, we investigate the wavelength dependence of the KMOS PSF. To do this, we use the reduced data cubes of stars targeted during the K-CLASH observations (see \citetalias{Tiley:2020} and Section \ref{sec:K-CLASH_survey}). We sum the flux in a window of width 0.05\,$\mu$m in the wavelength direction and fit a two-dimensional Gaussian function. This is repeated a further five times for increasing central wavelengths.  We plot the full-width at half-maximum (FWHM) of the best-fitting Gaussian against wavelength for each spectrograph in Figure \ref{fig:spectrographs_PSF_variation_with_lambda}. 

We find that the FWHM of the PSF improves by  $\approx$0\farcs1 from the low- to the high-wavelength end of the \textit{IZ} band spectral range, although this effect is negligible for the results of this study. We also find that the PSFs of the three spectrographs are very comparable across the \textit{IZ} band; the largest difference between spectrographs is only 0\farcs03 at 1.05 $\mu$m. 

\begin{figure}
\begin{center}
\includegraphics[width=0.5\textwidth]{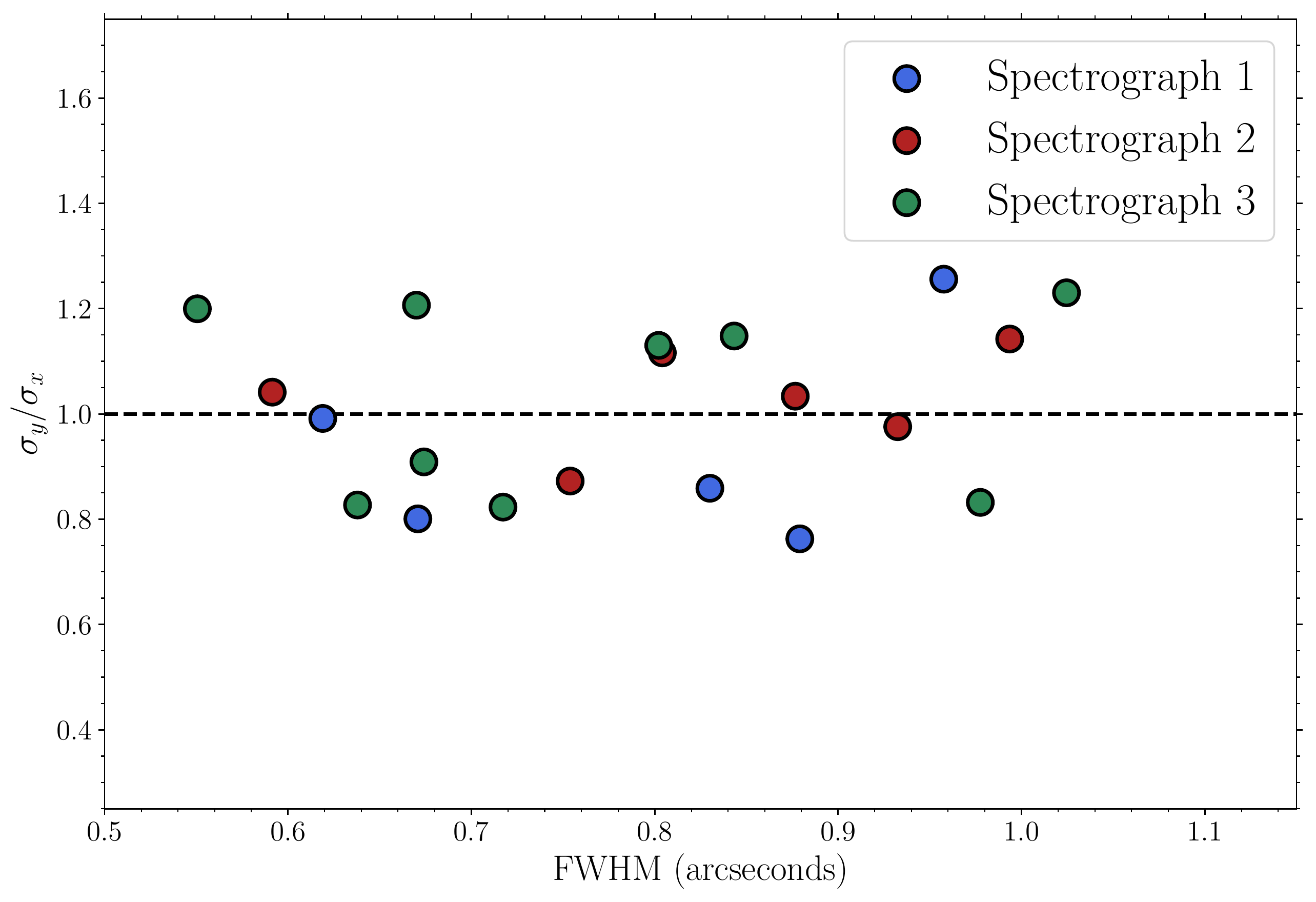}
\caption[Size and shape of the KMOS PSF as a function of spectrograph]{Ratio of the standard deviations ($\sigma_{x}$ and $\sigma_{y}$) of the two-dimensional Gaussian fit to each star observed during the K-CLASH survey, plotted against the FWHM of the two-dimensional Gaussian fit. Points are colour-coded according to the KMOS spectrograph they were observed with. Each PSF is generally circular, with deviation from circularity at worst $\approx25$\%.}
\label{fig:spectrographs_PSF_variation}
\end{center}
\end{figure}

We also compare all of the star observations taken during the K-CLASH survey to investigate the stability of the KMOS PSF over time, as well as any deviation from circularity. To do so, we collapse each datacube along the wavelength direction and fit the resulting image with a two-dimensional Gaussian. The ratio of the best-fitting dispersion values in the $x$ and $y$ directions ($\sigma_x$ and $\sigma_y$) for each star observation is shown in Figure \ref{fig:spectrographs_PSF_variation}, plotted against the FWHM of the best-fitting Gaussian. 

We find that the PSF is generally circular, with deviations from circularity of at worst $\approx$25\%. The average $\frac{\sigma_x}{\sigma_y}$ is 0.96, 1.04 and 1.06 for spectrographs 1, 2 and 3 respectively. Furthermore, we account for these small differences between spectrographs by using the PSF image observed with the same spectrograph as the science data whenever possible.

\section{\Ha{} Line Map Signal-to-Noise tests}
\label{sec:SN_tests}

To test the robustness of our \Ha{} spatial profile measurements, we create mock datacubes with model \Ha{} surface brightness distributions and fit them in exactly the same way as real observations. We take the mock radial \Ha{} surface brightness profiles to be those of exponential discs, using \imfit{} to create a two-dimensional surface brightness distributions. The disc models have six free parameters: 

\begin{itemize}
\item the coordinates of the image centre,  $x_0$ and $y_0$
\item the observed ellipticity, $\epsilon$, defined as $1- \frac{b}{a}$ (where $a$ and $b$ are respectively the semi-major and semi-minor axes of an ellipse)
\item the disc position angle, PA, measured counter-clockwise from the positive $y$ axis of the image
\item the central intensity, $I_0$
\item the exponential disc scale length, $R_{\rm{disc}}$\footnote{We note that in Section \ref{sec:Halpha_size} we have converted all measured values of exponential scale length into half-light radii, $r_{\rm{e, H}\alpha}$, by performing a curve of growth analysis in a circular aperture on the intrinsic (unconvolved) best-fitting model.}
\end{itemize}

\noindent We then create a datacube by assigning a mock spectrum to each pixel in the image. We model the continuum light as a second order polynomial, and superimpose a single Gaussian emission line at a wavelength corresponding to \Ha{} emission at $z=0.4$. This emission line template has velocity dispersion of 100 \kms{} and a peak flux equal to twice the continuum level. The absolute normalisation of each spectrum is defined by the value of the model surface brightness distribution at that spaxel. 

Random noise is then added. This is accomplished by taking the mean spectrum of the model cube from within its half light radius and dividing by an input S/N value to create a noise "spectrum". At each wavelength slice, random numbers are then drawn from a Gaussian distribution (centred on zero with width corresponding to the value in the noise spectrum) and added to the cube. This process implies that the \textit{average} S/N value of pixels within the model half-light radius of each datacube is equal to the requested S/N ratio; the S/N at the centre of the image and at the edges will be respectively higher and lower than this average. Finally, the mock cube is convolved with the KMOS PSF and the parameters of \Ha{} surface brightness distribution are measured in the manner discussed in Section \ref{sec:Halpha_size}. We note that these simulations do not model the effect of sky-subtraction residuals on our measurement process. 

\begin{figure}
\begin{center}
\includegraphics[width=0.5\textwidth]{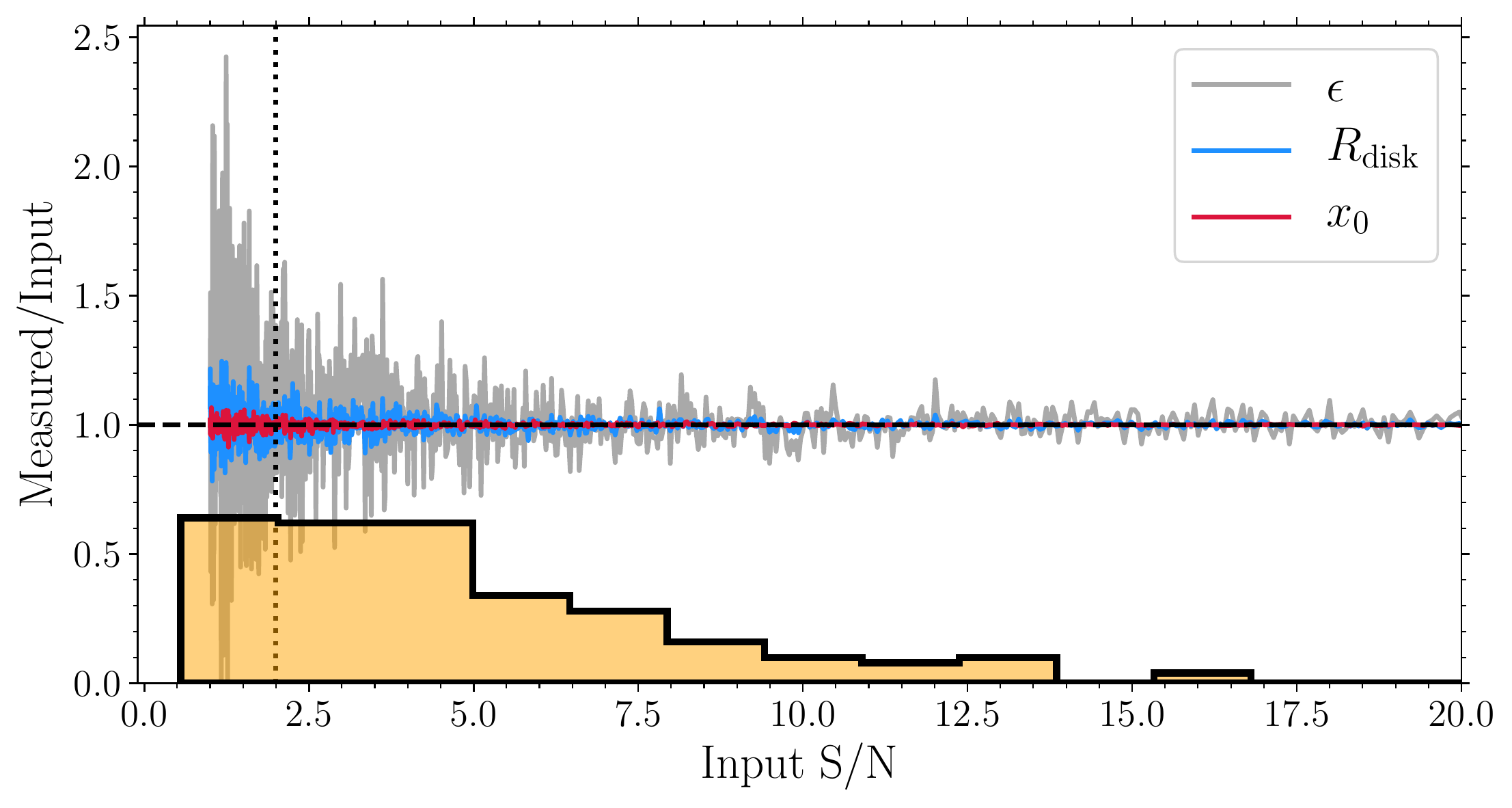}
\caption[Mock tests on \Ha{} image fitting I]{Recovery of the \Ha{} profile parameters of mock observations, at various input S/N ratios of the H$\alpha$ images. Note the image S/N value is the \textit{average} S/N value of pixels within the best-fitting half-light radius. The dashed line represents an average image S/N of 2, and the histogram shows the S/N distribution of our observed \Ha{} images described in Section \ref{sec:Ha_line_maps}. See Appendix \ref{sec:SN_tests} for details.}
\label{fig:SN_tests}
\end{center}
\end{figure}

 \begin{figure}
\begin{center}

\includegraphics[width=0.5\textwidth]{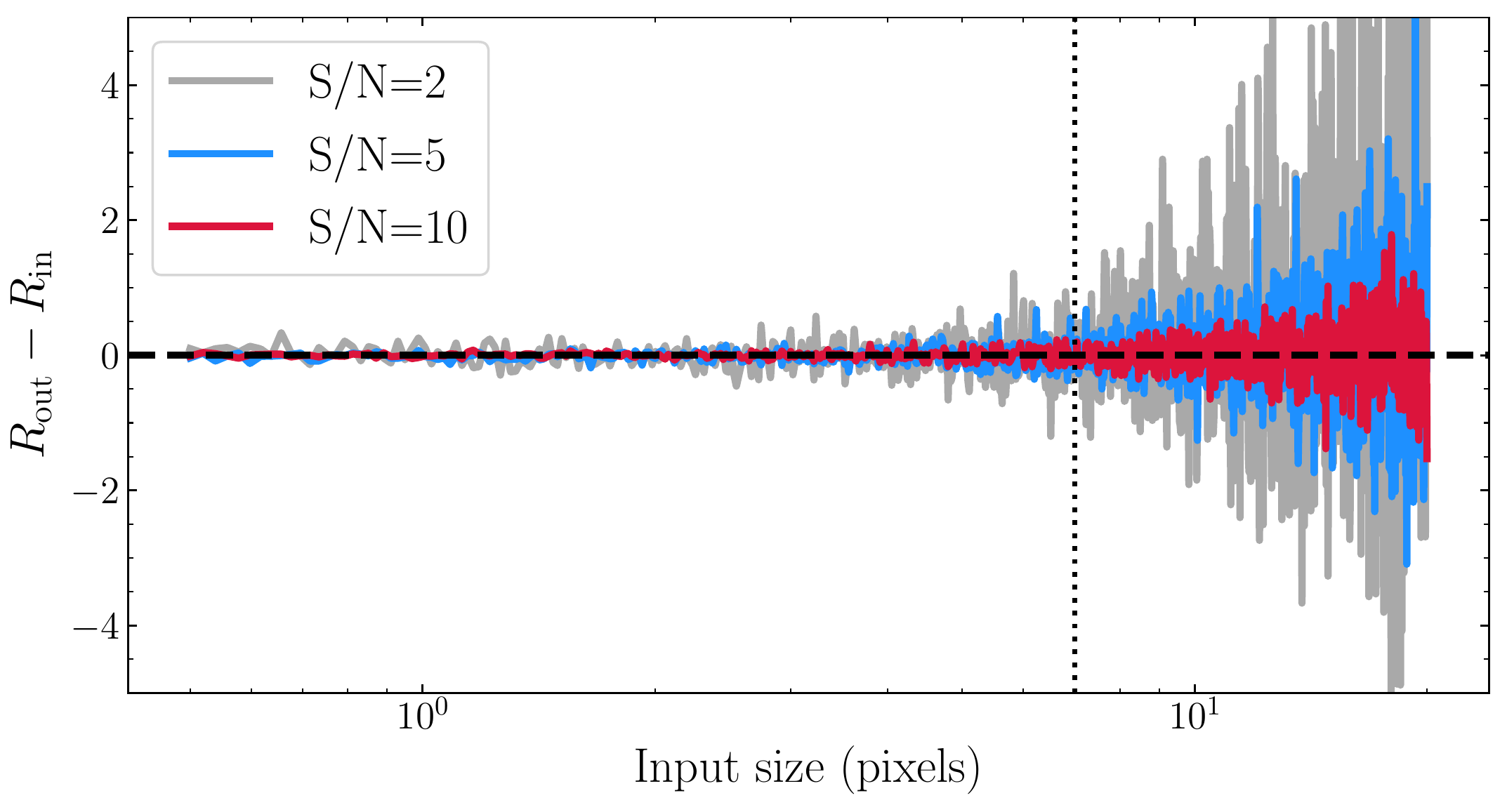}
\includegraphics[width=0.5\textwidth]{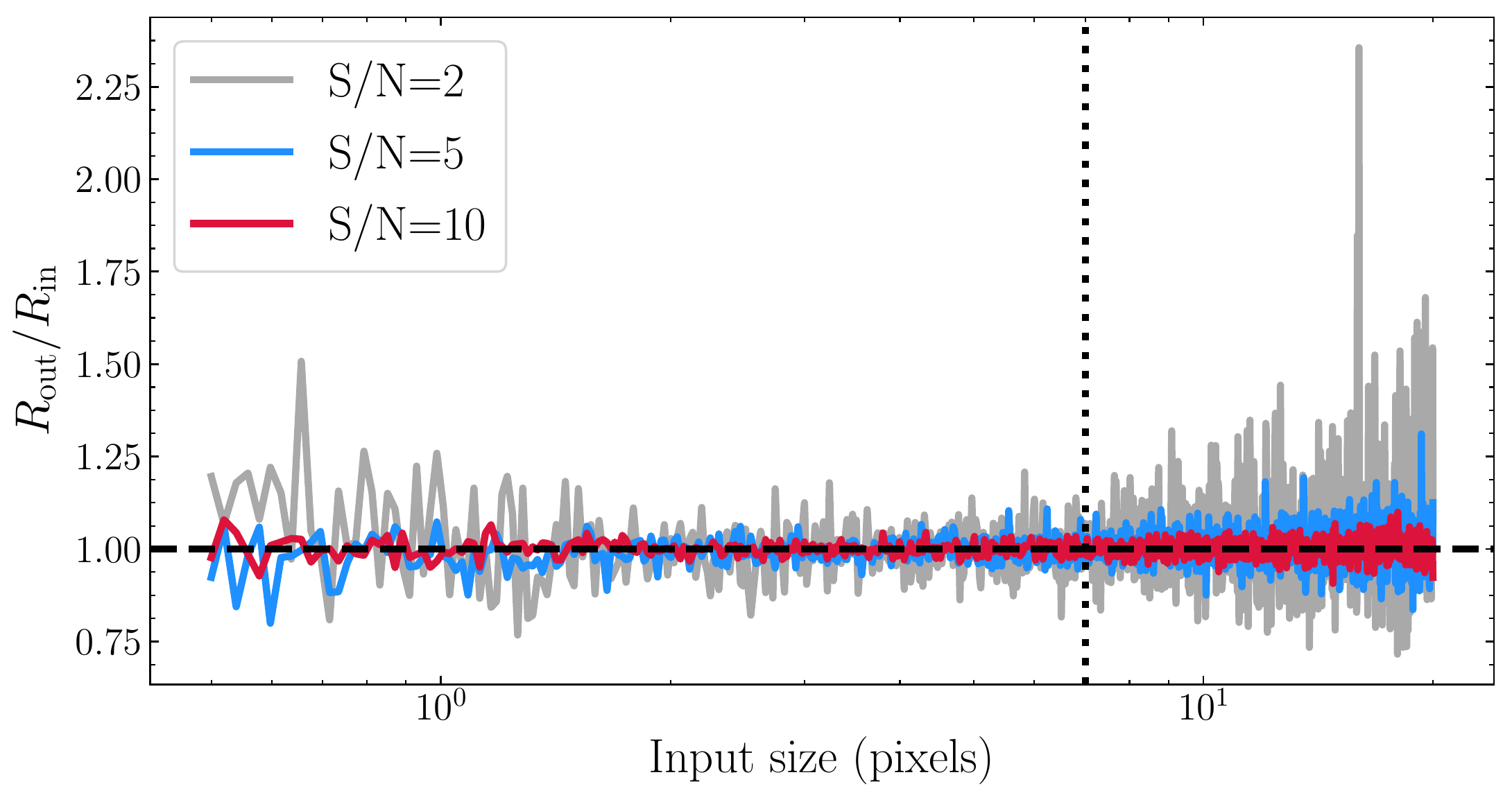}
\caption[Mock tests on \Ha{} image fitting II]{Recovering the \Ha{} disc scale length ($R_{\mathrm{disc}}$) of mock observations, for various scale lengths and S/N ratios. $R_{\rm{in}}$ refers to the true value of the model exponential scale length whilst $R_{\rm{out}}$ refers to the measured value. The dashed line represents the size of the KMOS IFU for a source placed in the centre (7 pixels). The top panel shows the difference between the input and output scale lengths, whilst the bottom shows their ratio.}
\label{fig:SN_tests_2}
\end{center}
\end{figure}

 \begin{figure}
\begin{center}
\includegraphics[width=0.5\textwidth]{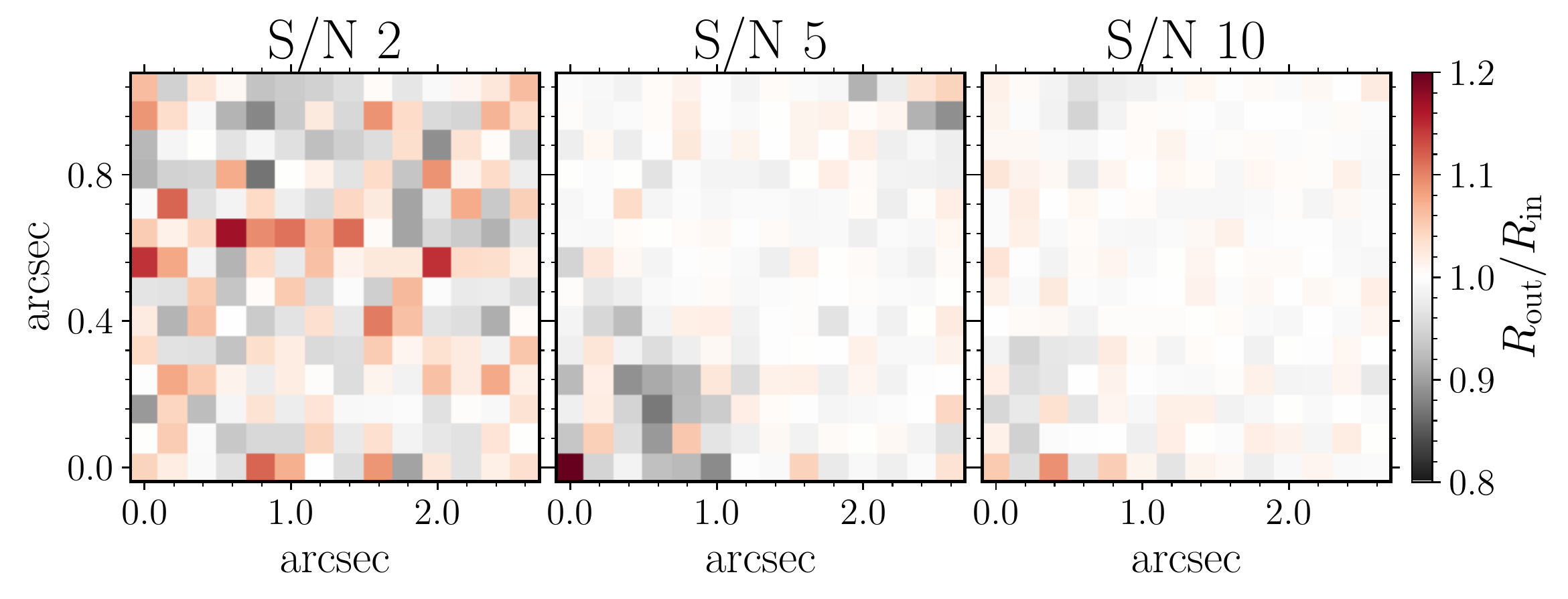}
\caption[Mock tests on \Ha{} image fitting III]{Recovering the \Ha{} disc scale length ($R_{\mathrm{disc}}$) of mock observations, for various locations on the IFU and S/N values. The model is a source with $R_{\rm{disc}}=3$ pixels, $\epsilon=0.2$ and position angle 135\textdegree.}
\label{fig:SN_tests_3}
\end{center}
\end{figure}

The results of our tests are shown in Figures \ref{fig:SN_tests}, \ref{fig:SN_tests_2} and \ref{fig:SN_tests_3}. Figure \ref{fig:SN_tests} shows that estimating $R_{\rm{disc}}$ requires a larger S/N than simply finding the ($x_0, y_0$) coordinate of the \Ha{} flux centre, but is easier than constraining the galaxy ellipticity ($\epsilon$). A histogram of the S/N ratios of our data is shown in orange. We find that to recover $R_{\rm{disc}}$ to an accuracy better than 10\%, we require the average of the S/N of the integrated \Ha{} flux in all spaxels within the best-fitting half-light radius to be greater than 2. For an exponential surface brightness profile, this implies that the central S/N ratio is $\approx10$. 

To investigate the effect of the finite size of each KMOS IFU, we placed a mock galaxy in the centre of an IFU and varied the input $R_{\rm{disc}}$, with a range of average S/N ratios. Figure \ref{fig:SN_tests_2} shows comparisons of the input ($R_{\rm{in}}$) and recovered ($R_{\rm{out}}$) disc scale lengths as a function of $R_{\rm{in}}$ and S/N. For reference, the average half-light radius at $z=0.5$ is approximately 1\arcsec \citep[4 kpc; ][]{Paulino-Afonso:2017}, which corresponds to a disc scale-length of $\approx$0\farcs6 ($\approx$ 3 pixels). We find that our ability to recover $R_{\rm{disc}}$ is good to better than 10\% for high S/N data at all values of $R_{\rm{in}}$ , showing that the limited size of the KMOS field of view does not hinder these measurements.  At S/N=2, we recover galaxies almost three times the average disc-scale length (11 pixels) with a 25-50\% uncertainty. 

Finally, we assess the impact of mis-centred \Ha{} emission by placing our mock galaxy at various positions across the IFU. We create a model galaxy with $\epsilon=2$, $R_{\rm{disc}}=3$ pixels and position angle 135\textdegree{} and then place it at various $(x, y)$ locations. We measure the best-fitting $R_{\rm{disc}}$ 10 times, and show the average ratio of $R_{\rm{out}}$/$R_{\rm{in}}$ for S/N of 2, 5 and 10 in Figure \ref{fig:SN_tests_3}. We find that small offsets from the centre have no effect on our ability to recover $R_{\rm{disc}}$. As expected, the largest uncertainties occur when the mis-centring is large (i.e. the object is in a corner of the IFU). We ensured this did not occur for any of our observations. 

Whilst not exhaustive, these tests show that we can robustly measure $R_{\rm{disc}}$ for a variety of S/N ratios, galaxy sizes and locations on the IFU, and allow us to make an informed decision on the minimum S/N ratio to use in our analyses. It should nevertheless be stressed that these tests are conducted under more favourable conditions than the real observations and analyses, since they are fitting a model which we know to be the true representation of the data, and do not include systematic uncertainties such as sky line residuals in the spectral dimension or "hot" pixels in the \Ha{} images.

\section{Table of Measurements}

We present all measurements used in this work in Table \ref{tab:all_KCLASH_values}. This table is available online in machine-readable format. 

\begin{table*}
\centering
\label{tab:all_KCLASH_values}
\caption{A sample of the measurements presented in this paper. The full table is available online in machine-readable format. \textbf{(1)} Right Ascension. \textbf{(2)} Declination. \textbf{(3)} The observation field the galaxy is located in. \textbf{(4)} Spectroscopic redshift from the \Ha{} line. \textbf{(5)} Do we detect \Ha{} in either the 0\farcs6, 1\farcs2 or 2\farcs4 aperture at a S/N ratio greater than 5? \textbf{(6)} Is the galaxy a member of the cluster sample? \textbf{(7)} Is the galaxy a member of the mass-matched field sample? \textbf{(8)} Does the galaxy contain an AGN? \textbf{(9)} Integrated flux in a 0\farcs6 diameter aperture. \textbf{(10)} Uncertainty on \textbf{(9)}. 
\textbf{(11)} $\log$ stellar mass derived from the SED fitting code Prospect (See \citetalias{Tiley:2020} for details).
\textbf{(12)} Velocity difference to cluster redshift (\kms). 
\textbf{(13)} Radius from cluster centre (arcseconds).
\textbf{(14)} Projected radius scaled to the cluster $R_{200}$ value.
\textbf{(15)} Half light radius from the \Ha{} image (arcseconds). We fit models to the galaxy \Ha{} surface brightness distribution, then measure $r_{\mathrm{e}}$ from a curve-of-growth analysis (integrating the best-fit model in a circular aperture;  See Section \ref{sec:Ha_surface_brightness_profiles}).
\textbf{(16)} Half light radius from the $R_c$ band image (arcseconds). We fit models to the galaxy image then measure $r_{\mathrm{e}}$ from a curve-of-growth analysis (integrating the best-fit model in a circular aperture; See Section \ref{sec:continuum_imaging}).
\textbf{(17)} Ratio of effective radii: \sizeratio.
\textbf{(18)} Does this galaxy meet the criteria described in Section 3.5 such that its size measurement is considered reliable?
\textbf{(19)} $\chi^2$ value from the \Ha{} image fitting.
\textbf{(20)} $\chi^2$ value from the $R_c$-band image fitting.
\textbf{(21)} Average S/N from within $r_{\mathrm{e}}$ in the \Ha{} image (see Section \ref{sec:Ha_line_maps}).
\textbf{(22)} Best-fit S\'ersic index from the $R_c$-band image.
\textbf{(23)} Best-fit S\'ersic index uncertainty.
\textbf{(24)} S/N of the \Ha{} line in a spectrum integrated in a 2\farcs4 diameter aperture. 
\textbf{(25)} \Ha{} flux measured by \texttt{ppxf} from a 2\farcs4 diameter aperture.
\textbf{(26)} \Ha{} flux uncertainty.
\textbf{(27)} [\ion{N}{ii}] flux measured by \texttt{ppxf} from a 2\farcs4 diameter aperture.
\textbf{(28)} [\ion{N}{ii}] flux uncertainty.
\textbf{(29)} Ratio of the [\ion{S}{ii}] lines. 
\textbf{(30)} [\ion{S}{ii}] ratio uncertainty. 
\textbf{(31)} Total [\ion{S}{ii}] flux measured by \texttt{ppxf} from a 2\farcs4 diameter aperture.
\textbf{(32)} Total [\ion{S}{ii}] flux uncertainty.
\textbf{(33)} $\chi^2$ from the \texttt{ppxf} fit. 
\textbf{(34)} Velocity dispersion from the \texttt{ppxf} fit (\kms). 
\textbf{(35)} Gas-phase metallicity ($12 + \log_{10} (\mathrm{O}/\mathrm{H})$) measured in a 2\farcs4 diameter aperture using the [\ion{N}{ii}]/\Ha{} ratio (see Section \ref{sec:emline_results}). 
\textbf{(36)} Gas-phase metallicity uncertainty.
}
\begin{tabular}{lcccccc}
\toprule
ID                                                                                      &           41309 &           41423 &           44143 &           45189 & 45677 &           46622\\
\midrule
RA \textbf{(1})                                                                         &         197.739 &         197.781 &         197.797 &         197.691&  197.795 &         197.678 \\
Dec        \textbf{(2})                                                                 &        -3.25943 &        -3.26086 &        -3.25042 &        -3.24392&   -3.243 &        -3.23713 \\
Observation Field      \textbf{(3})                                                         &        MACS1311 &        MACS1311 &        MACS1311 &        MACS1311 & MACS1311 &        MACS1311 \\
Spectroscopic Redshift \textbf{(4})                                                     &        0.271578 &         0.45068 &         0.36509 &        0.435555 &  0.48838 &        0.437338 \\
Detected H$\alpha$? \textbf{(5})                                                             &            True &           False &           False &            True    &    False &            True \\
Cluster sample  \textbf{(6})                                                            &           False &           False &           False &           False &    False &           False \\
Mass-matched field sample  \textbf{(7})                                                              &            True &           False &           False &            True &    False &            True \\
AGN flag     \textbf{(8})                                                               &           False &           False &           False &           False &    False &           False \\
$I_{\mathrm{H}\alpha}$ (D0.6; erg s$^{-1}$ arcsec$^{-1}$) \textbf{(9})                                              &      1.7911e-17 &     4.02604e-18 &               0 &     1.37713e-17  &        0 &     1.26111e-17 \\
$I_{\mathrm{H}\alpha}$ uncertainty (D0.6; erg s$^{-1}$ arcsec$^{-1}$)  \textbf{(10})                                 &     5.91279e-18 &     3.58005e-18 &               0 &     2.79143e-18   &        0 &     2.24848e-18 \\
$\log_{10} M_{*}$ $(M_{\odot})$ \textbf{(11})                                            &         10.0449 &          10.396 &         10.1432 &         10.3793 &  10.6962 &          10.173 \\
Cluster-centric velocity (\kms) \textbf{(12})                                           &        -44632.1 &        -8692.74 &        -25867.5 &        -11727.8   & -1127.82 &          -11370 \\
Projected radius (\arcsec) \textbf{(13})                                          &         301.678 &         310.833 &         297.111 &         337.239    &  270.619 &          355.75 \\
Projected radius/$R_{200}$ (\arcsec)   \textbf{(14})                                                  &         1.33089 &         1.37128 &         1.31074 &         1.48777    &  1.19386 &         1.56943 \\
H$\alpha$ $r_{\mathrm{e}}$ (\arcsec) \textbf{(15})                                                           &         1.59826 &            --- &            --- &        0.461888     &     --- &        0.728819 \\
Stellar $r_{\mathrm{e}}$ (\arcsec) \textbf{(16})                                                    &        0.582627 &        0.395013 &         1.07183 &        0.511791     & 0.309265 &        0.678999 \\
H$\alpha$-to-stellar size ratio  \textbf{(17})                                          &          2.7432 &            --- &            --- &        0.902494    &     --- &         1.07337 \\
Reliable size measurement \textbf{(18)}					&          True &            False &            False &        True    &     False &         True \\
$\chi^2$- H$\alpha$ image fitting  \textbf{(19})                                         &         3.13659 &            --- &            --- &          3.1118   &     --- &         2.42207 \\
$\chi^2$- stellar image fitting   \textbf{(20})                                         &         1.20319 &         1.34134 &         2.51276 &         2.38855   &  3.52322 &          2.9806 \\
Mean S/N (H$\alpha$ image fitting) \textbf{(21})                                         &         2.48416 &            --- &            --- &         3.69481    &     --- &         4.02895 \\
S\'ersic index (stars)     \textbf{(22})                                                 &          0.9406 &         2.73483 &         1.17899 &         1.14667  &   3.4329 &         0.56021 \\
S\'ersic index uncertainty (stars)    \textbf{(23})                                     &        0.031186 &        0.095936 &        0.017757 &        0.059112   &  0.14053 &         0.01166 \\
H$\alpha$ line S/N (D2.4)   \textbf{(24})                                               &         12.6412 &         4.37531 &               0 &               0  &        0 &         25.2812 \\
\texttt{ppxf} \Ha{} flux (D2.4; erg s$^{-1}$ arcsec$^{-1}$)  \textbf{(25})                      &     1.93485e-16 &            --- &            --- &     6.01279e-17   &     --- &     2.18856e-16 \\
\texttt{ppxf} \Ha{} flux uncertainty (D2.4; erg s$^{-1}$ arcsec$^{-1}$)  \textbf{(26})          &     2.10491e-13 &            --- &            --- &               0    &     --- &     7.85701e-18 \\
\texttt{ppxf} \ion{N}{ii} flux (D2.4; erg s$^{-1}$ arcsec$^{-1}$)  \textbf{(27})                    &     3.39716e-18 &            --- &            --- &     8.66973e-18   &     --- &     1.22466e-16 \\
\texttt{ppxf} \ion{N}{ii} flux uncertainty (D2.4; erg s$^{-1}$ arcsec$^{-1}$) \textbf{(28})     &     2.11108e-13 &            --- &            --- &               0    &     --- &     1.31339e-17 \\
\texttt{ppxf} \ion{S}{ii} ratio (D2.4)  \textbf{(29})                                           &           -- &            --- &            --- &            --   &     --- &            1.43 \\
\texttt{ppxf} \ion{S}{ii} ratio uncertainty (D2.4) \textbf{(30})                                    &        -- &            --- &            --- &           --    &     --- &       0.0967388 \\
\texttt{ppxf} \ion{S}{ii} flux total (D2.4; erg s$^{-1}$ arcsec$^{-1}$) \textbf{(31})           &               0 &            --- &            --- &     3.00684e-17    &     --- &     1.06618e-16 \\
\texttt{ppxf} \ion{S}{ii} flux total uncertainty (D2.4; erg s$^{-1}$ arcsec$^{-1}$) \textbf{(32})&     7.54642e-14 &            --- &            --- &               0     &     --- &     1.02338e-17 \\
\texttt{ppxf} $\chi^2$ (D2.4) \textbf{(33})                                                     &         1.87915 &            --- &            --- &          4.4041  &     --- &         5.10002 \\
\texttt{ppxf} velocity dispersion (D2.4; \kms) \textbf{(34})                                        &         68.8751 &            --- &            --- &         45.3285   &     --- &          100.05 \\
$12 + \log_{10} (\mathrm{O}/\mathrm{H})$   (D2.4) \textbf{(35})                                         &         7.88033 &            --- &            --- &        ---  &     --- &         8.76653 \\
$12 + \log_{10} (\mathrm{O}/\mathrm{H})$ uncertainty  (D2.4) \textbf{(36})                                  &         1.08415 &            --- &            --- &               ---   &     --- &       0.0601683 \\
\bottomrule
\end{tabular}

\end{table*}

\bsp	\label{lastpage}
\end{document}